# Thermoviscoelasticity of polydomain liquid crystal elastomers regulated by soft elasticity

Zhengxuan Wei[a,1], Beijun Shen[b,*,1], Zumrat Usmanova[a], Umme Hani Bootwala[a] and Ruobing Bai[a,*]

[a]Department of Mechanical and Industrial Engineering, College of Engineering, Northeastern University, Boston, MA, 02115, United States
[b]Department of Mechanical Engineering, Columbia University, New York, NY, 10027, United States

ARTICLE INFO

ABSTRACT

Liquid crystal elastomers (LCEs) are elastomeric networks with rod-like mesogens that reorient under load. In polydomain LCEs, this reorientation drives a polydomain-to-monodomain transition that produces a soft-elastic plateau. The intricate coupling between this soft elasticity and the polymer-network viscoelasticity gives rise to a path-dependent thermoviscoelastic response, central to applications of polydomain LCEs in damping, impact protection, and tough adhesives. However, the physics governing this response under complex thermomechanical histories remains insufficiently studied. In this work, we present a combined experimental and theoretical study of polydomain LCEs under three uniaxial loading protocols: single-cycle loading-unloading, stress-free recovery from various pre-stretches, and multi-cycle loading with progressively increasing amplitude. We develop a finite-deformation constitutive model that combines two parallel mechanisms: rate-independent, temperature-dependent soft elasticity from mesogen reorientation, and time- and temperature-dependent viscoelasticity from various sources. Calibrated with a single parameter set, the model quantitatively reproduces all three protocols and resolves the individual contribution of each mechanism. Across the three protocols, a temperature-dependent soft-elastic limit governs the low-rate response and the long-time recovered stretch, while polymer-network viscoelasticity controls the rate-dependent deviation and the cycle-wise accumulation of residual stretch away from this low-rate limit. A thermal recovery test above the nematic-isotropic transition temperature confirms that all hysteresis and residual deformation are reversible, ruling out irreversible internal damage. The combined experimental-theoretical framework provides mechanistic understanding and a predictive basis for the design of polydomain LCE components under complex thermomechanical histories.

## 1. Introduction

Liquid crystal elastomers (LCEs) are stimuli-responsive polymer networks of crosslinked rubbery chains with embedded rod-like mesogens. Most modern LCEs are thermotropic: mesogens align in a nematic phase at low temperatures and disorder into an isotropic phase above a nematic-isotropic transition temperature $T_{ni}$, typically around 60 °C (Kularatne et al., 2017; Traugutt et al., 2017; Wei et al., 2025). The coupling between mesogen alignment and polymer-network deformation gives LCEs distinctive thermomechanical properties that depend strongly on the underlying mesogen organization (Warner and Terentjev, 2007; Ware et al., 2015; White and Broer, 2015; Guin et al., 2018; Hertlein et al., 2023). *Monodomain* LCEs with a uniform mesogen alignment exhibit large, reversible thermal actuation, motivating their use in robotic actuators (Kularatne et al., 2017; Ula et al., 2018; Liu et al., 2021; Saed et al., 2022; Chen et al., 2024b) and shape-morphing structures (Sydney Gladman et al., 2016; Aharoni et al., 2018; Li et al., 2021; Wu et al., 2021; Li et al., 2022). *Polydomain* LCEs, the focus of the present work, instead consist of co-existing microscale nematic domains with different orientations and are macroscopically isotropic (Fig. 1a). These materials do not undergo direct thermal actuation (Wei et al., 2025), but exhibit other distinct mechanical responses arising from their microstructure, such as soft elasticity and significant viscoelasticity detailed in the following.

In a polydomain LCE, under tensile load in an arbitrary direction, mesogens in different domains progressively reorient toward the loading direction. The material undergoes large deformation (often > 50% engineering strain at room temperature) with a near-zero plateau stress, a phenomenon known as *soft elasticity* (Conti et al., 2002a; Warner and Terentjev, 2007; Urayama et al., 2009; Biggins et al., 2009, 2012; Bai and Bhattacharya, 2020; Tokumoto et al.,

*Corresponding author
bs3660@columbia.edu (B. Shen); ru.bai@northeastern.edu (R. Bai)
[1]These authors contributed equally to this work.

2021; Wei et al., 2025) (Fig. 1b). Once mesogen reorientation is complete, the LCE is effectively monodomain and stiffens under further stretch like a conventional rubber.

The mesogen-polymer coupling also gives polydomain LCEs pronounced viscoelasticity. Beyond classical mechanisms such as reptation and intermolecular friction of polymer chains (Ferry and Rice, 1962; Rubinstein and Colby, 2003), deformation-induced mesogen reorientation contributes to strongly rate- and temperature-dependent viscoelastic dissipation in LCEs (Azoug et al., 2016; Luo et al., 2021; Wei et al., 2023a; Zhou et al., 2025). This viscoelasticity acts synergistically with soft elasticity to enable emerging applications of polydomain LCEs, including damping and impact protection (Linares et al., 2020; Traugutt et al., 2020; Jeon et al., 2022; Shen, 2025; Shen et al., 2026), tough adhesives (Annapooranan et al., 2024; Choi et al., 2025; Koshimizu and Saed, 2025), biologically compliant materials (Ware et al., 2015), and architected energy-absorbing metamaterials (Fu et al., 2019; Liang and Li, 2022; Jeon et al., 2022; Shen, 2025; Shen et al., 2026). Furthermore, temperature plays a pivotal role in determining the rate- and history-dependent mechanical response, since both mesogen reorientation and polymer-network relaxation are intrinsically temperature-dependent.

The expanding application space of polydomain LCEs have spurred experimental investigations into their thermoviscoelastic responses. Dynamic mechanical analysis (DMA) and stress relaxation tests revealed multiple relaxation timescales in LCEs associated with mesogen reorientation and polymer-network relaxation (Hotta and Terentjev, 2003). Subsequent DMA and tensile tests across temperatures and strain rates showed that the polydomain-to-monodomain transition is strongly rate-dependent and follows time-temperature superposition (TTS) (Azoug et al., 2016). Uniaxial tensile tests of main-chain nematic LCEs unveiled two well-separated relaxation timescales (Luo et al., 2021; Wei et al., 2023a; Zhou et al., 2025): a shorter one associated with mesogen reorientation and a longer one with polymer-network relaxation, each contributing measurably to the rate-dependent stress-stretch response. Under cyclic loading, polydomain LCEs exhibit large stress-stretch hysteresis at room temperature (Traugutt et al., 2020; Wu et al., 2021; Dai et al., 2025), and this dissipation has been amplified through architected structures that combine viscoelasticity with structural instability (Jeon et al., 2022; Shen, 2025; Shen et al., 2026).

The experimental progress has been paralleled by the active development of theoretical models for thermoviscoelastic LCEs. The pioneering neo-classical hyperelastic model of Bladon et al. (1993) introduced a step-length tensor to the free energy that captures soft elasticity in nematic LCEs. Verwey and Warner (1997) extended the framework to incorporate viscous director rotation in describing rate-dependent behavior. For monodomain LCEs, Wang et al. (2022) developed a nonlinear finite-deformation viscoelastic micropolar theory that decouples the viscous contribution of mesogen reorientation from polymer-network relaxation, capturing the rate-dependent stress-stretch response at room temperature. Similar framework has been introduced to study fracture of LCEs (Wei et al., 2024) and general three-dimensional finite element analysis (Chehade et al., 2024). Jiang et al. (2026) developed a viscoelastic micro-stretch theory for monodomain LCEs that captures thermal deformation across the nematic-isotropic transition. Multi-relaxation viscoelastic descriptions have also been developed for monodomain LCEs with exchangeable disulfide bonds, capturing anisotropic rate-dependent response at room temperature (Chen et al., 2025). Complementary efforts on monodomain LCEs have extended the rate range across six decades from quasi-static to dynamic loading, by combining director and order-parameter evolution with continuum-level viscoelasticity (Chung et al., 2024), in monotonic loading at room temperature.

Compared to monodomain LCEs, thermoviscoelastic modeling of polydomain LCEs has been less investigated in limited loading conditions, stretch rates, and temperatures. Lee et al. (2023) introduced a macroscopic constitutive model for isotropic-genesis polydomain LCEs based on internal state variables for domain-pattern evolution, validated under multi-axial monotonic loading at room temperature. Building on this framework, Wihardja et al. (2026) added network viscoelasticity to extend the rate range from quasi-static to dynamic loading in tension and compression, in monotonic loading at room temperature. Kutsyy et al. (2025) adopted this framework to characterize the soft-elastic response across temperatures and strain rates, with model parameters fitted at each temperature rather than within a unified temperature-dependent description. Other efforts have addressed polydomain LCEs through viscoelastic models calibrated for high-strain-rate compression in impact applications (Wang et al., 2024), and through quasi-static thermomechanical coupling for temperature-driven actuation and polydomain-monodomain transition (Chen et al., 2024a, 2026), both focusing on monotonic loading.

Despite this rapid progress in both experiment and theory, two outstanding challenges remain in the fundamental understanding of thermoviscoelastic polydomain LCEs. Experimentally, there is still a lack of systematic experimental data spanning a broad range of temperatures, stretch rates, and history-dependent loading protocols, particularly those beyond monotonic uniaxial tension, such as single-cycle and multi-cycle loading-unloading, stress relaxation, and creep. These rheological tests have been extensively employed for other conventional elastomers (Bergström and Boyce, 1998; Amin et al., 2006), but have remained less complete in LCEs. Theoretically, a unified finite-deformation thermoviscoelastic framework that captures the path-dependent response of polydomain LCEs under these experimental protocols with a single set of time- and temperature-dependent parameters has so far received less attention.

Motivated by these gaps, here we combine systematic experiments and finite-deformation constitutive modeling to investigate the path-dependent thermoviscoelasticity of polydomain LCEs across temperatures, stretch rates, and loading histories. Section 2 describes the material synthesis and experimental protocols. Section 3 introduces a constitutive framework that combines two parallel mechanisms: a rate-independent, temperature-dependent soft-elastic mechanism for mesogen reorientation, and a nonlinear viscoelastic mechanism for the time- and temperature-dependent relaxation of the polymer network. The model is calibrated with a single set of parameters and used to predict experimental results under three uniaxial loading protocols at various temperatures and rates: single-cycle loading-unloading (Section 4), stress-free recovery (Section 5), and multi-cycle loading-unloading (Section 6). Across the three protocols, a temperature-dependent soft-elastic limit emerges as the unifying physical quantity governing the low-rate response and the long-time recovered stretch, while polymer-network viscoelasticity controls the rate-dependent deviation and the cycle-wise accumulation of residual stretch away from this low-rate limit. A thermal recovery test confirms that all observed hysteresis and residual deformation are reversible, ruling out irreversible internal damage. Section 7 summarizes the findings and outlines directions for future work.

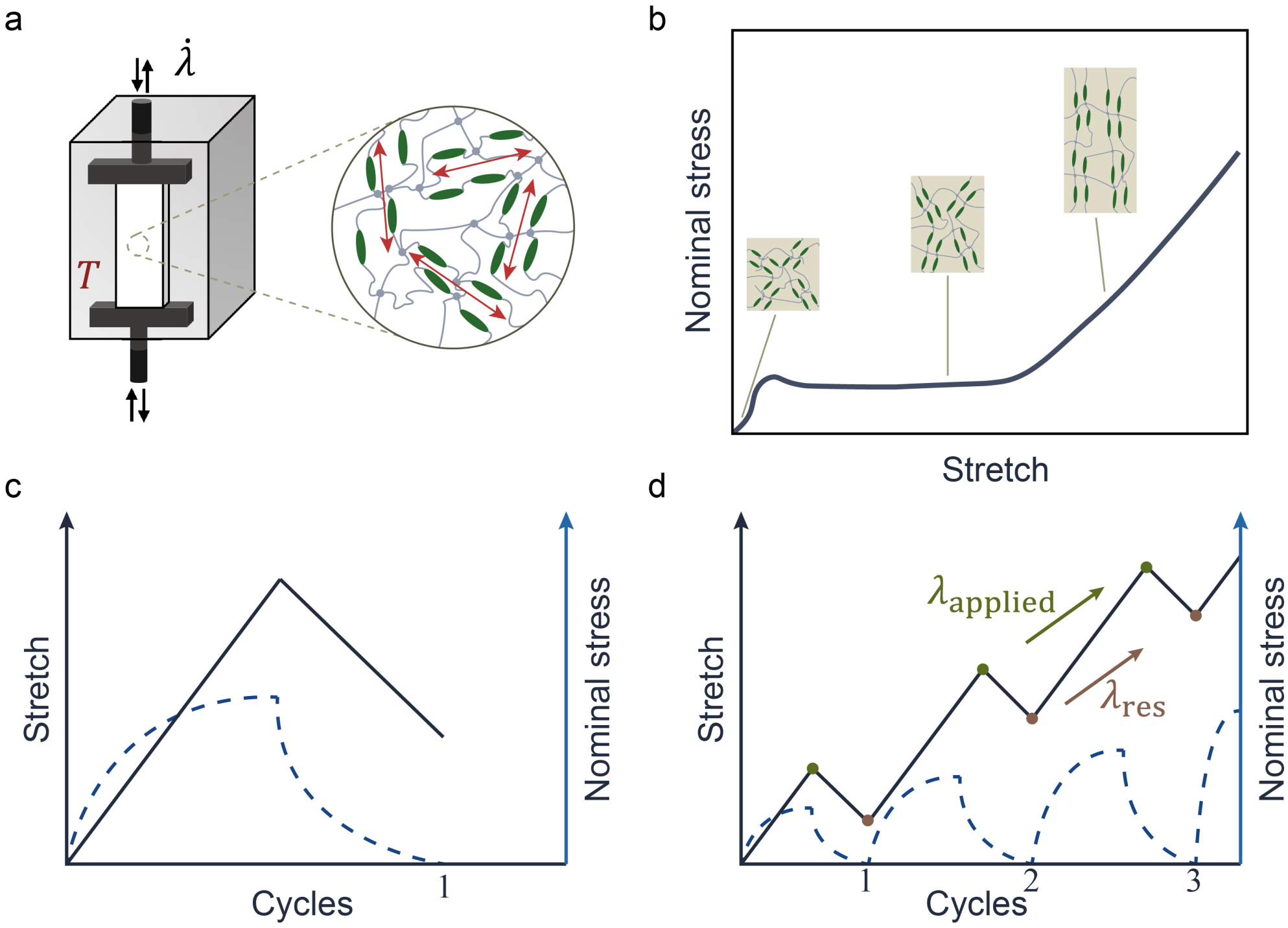

Figure 1: Schematics of experimental setup and uniaxial loading protocols at various temperatures and rates. (a) A tensile tester equipped with an environmental chamber, together with the molecular structure of the polydomain LCE. Green rods represent liquid crystal mesogens, gray lines and dots represent the polymer network, and red arrows represent the mesogen orientation in each domain. (b) A representative uniaxial stress-stretch response of polydomain LCE in the nematic phase with the corresponding evolution of mesogen reorientation toward the loading direction. (c) Single-cycle loading-unloading protocol at prescribed temperature and stretch rate. (d) Multi-cycle loading-unloading protocol at prescribed temperature and stretch rate, together with the cycle-wise applied maximum stretch ($\lambda_{\text{applied}}$, green dots) and recorded residual stretch ($\lambda_{\text{res}}$, brown dots). In (c)&(d), solid lines denote the applied stretch history and dashed lines denote the measured stress response. Each cycle ends with the first instance of zero nominal stress during unloading, at which point the cycle-wise residual stretch $\lambda_{\text{res}}$ is recorded. The next cycle begins immediately from this residual stretch with no dwell time between.

# 2. Experimental method

## 2.1. Synthesis of polydomain liquid crystal elastomer

4-bis-[4-(3-acryloyloxypropypropyloxy) benzoyloxy]-2-methylbenzene (RM257, mesogen) was purchased from Daken Chemical. Pentaerythritol tetra(3-mercaptopropionate) (PETMP, crosslinker), 2,2-(ethylenedioxy)diethanethiol (EDDET, spacer), 2,6-di-tert-butyl-4-methylphenol (BHT, antioxidant), 2-hydroxy-4-(2-hydroxyethoxy)-2-methylpropiophenone (HHMP, photoinitiator), dipropylamine (DPA, catalyst), and toluene (solvent) were purchased from Sigma-Aldrich.

Isotropic-genesis polydomain LCE samples are synthesized following the widely used two-stage polymerization method (Yakacki et al., 2015; Saed et al., 2016; Wei et al., 2025). We first form a precursor solution by mixing RM257 (8 g), BHT (0.16 g), toluene (3.2 g, 40 wt% of RM257), PETMP (0.434 g), HHMP (0.0514 g), and EDDET (1.83 g). The antioxidant BHT is added to the precursor to absorb extra undesired free radicals and increase the polymerization time (Traugutt et al., 2017). The mixture is heated to 85 °C and stirred continuously until a homogeneous solution is formed. Meanwhile, we prepare a catalyst solution of 1.136 g, with a weight ratio of DPA:toluene of 1:50. The catalyst solution is subsequently mixed with the precursor solution and degassed for 1.5 minutes using a FlackTek SpeedMixer. The mixture is poured into acrylic molds of $5\times1\times0.1$ cm$^3$ for polymerization. After 12 hours, the samples are taken out and placed in a vacuum oven at 80 °C and 508 mmHg for another 24 hours to evaporate the toluene. Following toluene evaporation, the polymerized LCE sample contracts to a length of approximately 4.5–4.7 cm and a final thickness of approximately 1–1.3 mm (the thickness can slightly exceed the depth of the 1 mm mold due to minor overfilling of the precursor solution).

## 2.2. Single-cycle and multi-cycle uniaxial tests at various temperatures and stretch rates

Single-cycle and multi-cycle uniaxial tests are performed using an Instron tensile tester (Instron 34TM-5) equipped with an environmental chamber that maintains a prescribed temperature with a real-time thermometer (Fig. 1a). A synthesized rectangular sample is mounted between two grips of the tensile tester, forming a fixed gauge length of 4 cm. The width and thickness of each sample are measured individually using a caliper immediately prior to testing to account for sample-to-sample variability, with typical values of approximately 9.4–10 mm in width and 1–1.3 mm in thickness, corresponding to an aspect ratio of about 4:1. The uniaxial nominal stress is calculated as the measured force divided by the cross-sectional area of the undeformed sample. The uniaxial stretch, denoted as $\lambda$, is calculated as the deformed length divided by the initial gauge length of the undeformed sample. Single-cycle loading-unloading tests are performed at three prescribed temperatures: 22 °C (room temperature, below $T_{ni}$), 40 °C (above room temperature, below $T_{ni}$), and 80 °C (above $T_{ni}$). Multi-cycle loading-unloading tests are performed at 22 °C and 40 °C, due to the near-zero hysteresis and the largely decreased ultimate stretch at 80 °C. Prior to each test, the sample is held at the prescribed temperature inside the environmental chamber for 3 minutes to ensure thermal equilibrium, which is also examined by observing the color change of the polydomain sample.

At each prescribed temperature, single-cycle tests are further performed at three different stretch rates of 0.1, 0.01, and 0.001 s$^{-1}$, spanning two orders of magnitude to probe the rate-dependent viscoelastic response. In each test, a sample is loaded from the undeformed reference state ($\lambda = 1$) to a prescribed maximum stretch and then unloaded back to $\lambda = 1$ at the same rate (Fig. 1c). The maximum stretch is set to $\lambda_{\text{applied}} = 2.5$, 2.0, and 1.5 at $T = 22$, 40, and 80 °C, respectively, to ensure that the deformation extends beyond the soft-elastic plateau at 22 and 40 °C and probes a sufficient rubber-like response at 80 °C.

At each prescribed temperature, multi-cycle tests are performed at two different stretch rates of 0.01 and 0.001 s$^{-1}$. The testing protocol employs a progressive loading scheme in which the maximum applied stretch $\lambda_{\text{applied}}$ is incrementally increased over successive cycles. Each cycle ends with the first instance of zero nominal stress during unloading, at which point the cycle-wise residual stretch $\lambda_{\text{res}}$ is recorded. The next cycle begins immediately from this residual stretch with no dwell time between, as illustrated in Fig. 1d. At 22 °C, six successive cycles are conducted with a sequence of prescribed maximum stretch $\lambda_{\text{applied}} = \{1.25, 1.5, 1.75, 2.0, 2.25, 2.5\}$. At 40 °C, four successive cycles are conducted with $\lambda_{\text{applied}} = \{1.25, 1.5, 1.75, 2.0\}$.

Similar single-cycle and multi-cycle tests are performed in the thermal recovery test, with the prescribed maximum stretch up to $\lambda_{\text{applied}} = 3$ at 0.01 s$^{-1}$ and 22 °C (Fig. 11). The multi-cycle tests here include 16 cycles with $\lambda_{\text{applied}}$ increased by 0.125 after each cycle until $\lambda_{\text{applied}} = 3$.

### 2.3. Stress-free recovery test

Stress-free recovery tests are performed to characterize the time- and temperature-dependent recovery of polydomain LCEs from a prescribed uniaxial pre-stretch (Fig. 5a). Rectangular samples are cut from the same polymerized LCE sheets to form an aspect ratio of at least 4:1. To more accurately measure the local stretch at the center of the sample, we draw periodic markers on the sample surface prior to testing. The sample is then placed on a hot plate that maintains a prescribed temperature, with the sample temperature confirmed by a separate thermometer placed on its top surface. The tests are conducted at the same three temperatures in the uniaxial stress-stretch tests: 22 °C, 40 °C, and 80 °C.

In each test, an initially undeformed sample is rapidly stretched to a prescribed pre-stretch $\lambda = \lambda_0$, released immediately, and then allowed to recover in a stress-free state on the hot plate. The time evolution of the recovered stretch $\lambda_{\mathrm{rec}} = \lambda(t)$ is tracked by the spacing between the surface markers using a video camera. The manual loading and release protocol does not precisely control the loading rate, unloading rate, or the dwell time at $\lambda_0$, leading to test-to-test scatter that is discussed further in Section 5. For each prescribed temperature, the applied pre-stretches span the temperature-dependent soft-elastic limit $\lambda_{SL}(T)$ measured from the single-cycle tests in Section 4: $\lambda_0 = 1.5$, 1.6, 1.8, 2.25, 2.5, and 3.0 at 22 °C, and $\lambda_0 = 1.3$, 1.5, 1.8, 2.0, 2.25, 2.5, and 3.0 at 40 °C and 80 °C.

The total duration of the video recording depends on the prescribed temperature. At 40 °C and 80 °C, the recovered stretch reaches a plateau in approximately 60 seconds, and the recording is terminated once the plateau is established. At 22 °C, recovery could proceed over a substantially longer timescale of up to 3 days. In those cases, the early-time evolution is captured by continuous video recording, and the long-term recovered stretch is subsequently measured manually with a ruler at intermittent times until the plateau is reached.

## 3. Theoretical model

### 3.1. Modeling framework and kinematics

The constitutive framework captures two coupled mechanisms in thermoviscoelastic polydomain LCEs: the temperature-dependent soft-elastic response from mesogen reorientation, and the rate- and temperature-dependent viscoelastic response of the polymer network. Correspondingly, the framework consists of two branches connected in parallel (Fig. 2). A *soft-elastic branch* (Branch A) is rate-independent yet temperature-dependent, representing the network elasticity in coupling with mesogen reorientation. A *viscoelastic branch* (Branch B) is Maxwell-like, comprising a nonlinear spring in series with a nonlinear viscous dashpot, representing the rate- and temperature-dependent viscoelasticity of various mechanisms.

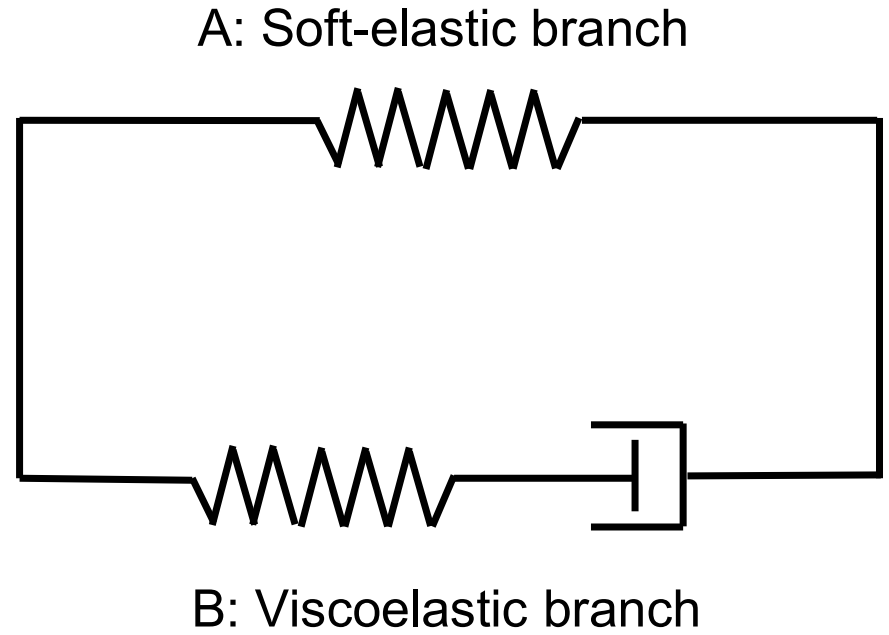

Figure 2: Schematic of the thermoviscoelastic constitutive framework. The total response of a polydomain LCE is represented by two branches in parallel. (A) A *soft-elastic branch* represents the rate-independent but temperature-dependent network elasticity in coupling with mesogen reorientation. (B) A *viscoelastic branch* represents the rate- and temperature-dependent viscoelasticity, comprising a nonlinear spring in series with a nonlinear viscous dashpot.

In the undeformed reference state, a polydomain LCE is macroscopically isotropic: co-existing nematic domains with distinct director orientations produce no globally preferred direction. Under tensile stretch, the sub-domain mesogens rotate cooperatively toward the loading direction. This physical process is captured by Branch A assuming ideal soft elasticity: the mesogen rotation accommodates the imposed stretch without storing elastic energy in the network, producing a zero-stress soft-elastic plateau in the stress-stretch curve of Branch A, until reaching a polydomain-monodomain transition point where the rotation is complete. On top of the ideal soft elasticity, Branch B provides additional viscoelastic dissipation of various mechanisms, including polymer-network relaxation, viscous mesogen rotation, sub-domain reorganization, and other contributions, all lumped into the nonlinear Maxwell-like element. This division of labor makes the modeling framework compact enough for calibration against single-cycle uniaxial stress-stretch tests (Section 4), yet predictive enough to reproduce the stress-free recovery (Section 5) and multi-cycle responses (Section 6) without recalibration.

Let $\chi(\mathbf{X}, t)$ denote the motion of a material point from the reference configuration $\mathbf{X}$ to the current configuration $\mathbf{x} = \chi(\mathbf{X}, t)$. The deformation gradient is

$$\mathbf{F} = \frac{\partial \chi}{\partial \mathbf{X}}, \tag{1}$$

and the total first Piola-Kirchhoff (nominal) stress is written as

$$\mathbf{P} = \mathbf{P}^{(A)} + \mathbf{P}^{(B)}, \tag{2}$$

where $\mathbf{P}^{(A)}$ and $\mathbf{P}^{(B)}$ are contributions from the soft-elastic and viscoelastic branches, respectively. For simplicity and consistency with most large-deformation constitutive treatments of LCEs, the material is taken to be incompressible, so that

$$\det \mathbf{F} = 1. \tag{3}$$

### 3.2. Soft-elastic branch

For the spring in Branch A, we adopt the quasi-convex elastic energy widely used for polydomain LCEs. This approach effectively treats the polydomain LCE as a homogenized continuum without resolving subdomain processes. Its free energy originates from the neo-classical theory of Bladon et al. (1993) and has been developed in subsequent work (Conti et al., 2002a,b; DeSimone and Dolzmann, 2002; He et al., 2020; Lee et al., 2023; Maghsoodi et al., 2023; Ahmadi and Maghsoodi, 2024). Our recent extension (Wei et al., 2023b; Usmanova and Bai, 2026) ensures a consistent description of temperature dependence between the underlying statistical model and the continuum form.

Following (Lee et al., 2023) and our recent extension (Usmanova and Bai, 2026), we take a Gent-like free-energy density (in unit reference volume) of Branch A as

$$\psi^{(A)} = \begin{cases} 0, & \text{Phase SE}: \lambda_1 \geq r_{\mathrm{LCE}}^{1/6} \\ -\dfrac{\mu_A(T)\, J_m\, r_{\mathrm{LCE}}^{-1/3}}{2(1-Q)} \ln\left[1 - \dfrac{\bar{I}^{(A)} - 3}{J_m}\right], & \text{else} \end{cases} \tag{4}$$

where $\bar{I}^{(A)}(\lambda_i, T)$ is the first invariant of the effective right Cauchy-Green tensor:

$$\bar{I}^{(A)}(\lambda_i, T) = \begin{cases} r_{\mathrm{LCE}}^{1/3}\left(\lambda_3^2 r_{\mathrm{LCE}}^{-1} + \lambda_2^2 + \lambda_1^2\right), & \text{Phase M}: r_{\mathrm{LCE}}^{-1/2}\lambda_3^2\lambda_1 > 1 \\ r_{\mathrm{LCE}}^{1/3}\left(\lambda_1^2 + 2\lambda_1^{-1} r_{\mathrm{LCE}}^{-1/2}\right). & \text{Phase P}: \text{else} \end{cases} \tag{5}$$

In Eqs. 4&5, $\mu_A(T)$ is the shear modulus of Branch A, $T$ is the absolute temperature, $J_m$ is the limiting-extensibility parameter, and $\lambda_1$, $\lambda_2$, and $\lambda_3$ are the three principal stretches that satisfy $\lambda_3 \geq \lambda_2 \geq \lambda_1$, with $\lambda_1\lambda_2\lambda_3 = 1$ assuming material incompressibility. $r_{\mathrm{LCE}}$ represents the temperature-dependent chain-anisotropy ratio, expressed as

$$r_{\mathrm{LCE}}(T) = \frac{1 + 2Q(T)}{1 - Q(T)}, \tag{6}$$

where $Q(T)$ is the temperature-dependent nematic order parameter. Furthermore, based on entropic elasticity, we assume $\mu_A(T)$ to be a linear function of $T$:

$$\mu_A(T) = \mu_{A0}\frac{T}{T_0}, \tag{7}$$

where $T_0 = 295$ K is the reference (room) temperature and $\mu_{A0} = \mu_A(T_0)$ is the modulus at $T_0$.

Eqs. 4&5 also define three effective *mechanical phases* that describe the characteristics of their associated mechanical responses (Conti et al., 2002a,b; DeSimone and Dolzmann, 2002; Lee et al., 2023; Wei et al., 2023b; Usmanova and Bai, 2026). The physical meanings and discussions of these three phases have been detailed in our recent work (Usmanova and Bai, 2026). Briefly, Phase SE (soft elastic) represents the ideal soft-elastic behavior, Phase P (planar) represents a liquid-like in-plane but solid-like out-of-plane behavior, and Phase M (monodomain) represents a conventional rubber-like behavior where all mesogens are aligned in a uniform direction.

For an incompressible hyperelastic material, the first Piola-Kirchhoff (nominal) stress in Branch A is given by

$$\mathbf{P}^{(A)} = \frac{\partial \psi^{(A)}}{\partial \mathbf{F}} - p^{(A)}\,\mathbf{F}^{-\mathrm{T}}, \tag{8}$$

where $p^{(A)}$ is the Lagrange multiplier determined by solving the boundary value problem.

### 3.3. Viscoelastic branch

Following Section 3.1, the kinematics of Branch B is described through the multiplicative decomposition of the deformation gradient

$$\mathbf{F} = \mathbf{F}_e \mathbf{F}_v, \tag{9}$$

where $\mathbf{F}_e$ and $\mathbf{F}_v$ are the elastic and viscous parts, respectively. The associated elastic left Cauchy-Green tensor is

$$\mathbf{B}_e = \mathbf{F}_e \mathbf{F}_e^{\mathrm{T}}, \tag{10}$$

and the total rate-of-deformation tensor is $\mathbf{D} = \mathrm{sym}(\dot{\mathbf{F}}\mathbf{F}^{-1})$.

We model the elastic spring in Branch B as an incompressible Arruda-Boyce eight-chain network (Arruda and Boyce, 1993). Introducing the elastic volume ratio $J_e = \det \mathbf{F}_e$ and the isochoric elastic left Cauchy-Green tensor $\bar{\mathbf{B}}_e = J_e^{-2/3}\mathbf{B}_e$, the elastic chain stretch is

$$\lambda_{\mathrm{chain},e} = \sqrt{\frac{1}{3}\mathrm{tr}\,\bar{\mathbf{B}}_e}. \tag{11}$$

The normalized chain-level force is expressed as

$$\beta_{\mathrm{chain},e} = \mathcal{L}^{-1}\left(\frac{\lambda_{\mathrm{chain},e}}{\sqrt{N}}\right), \tag{12}$$

where $\mathcal{L}^{-1}(\cdot)$ is the inverse Langevin function and $N$ is the number of Kuhn segments per chain. The corresponding Helmholtz free energy density is written as

$$\psi^{(B)} = \mu_B N \left[\frac{\lambda_{\mathrm{chain},e}}{\sqrt{N}}\,\beta_{\mathrm{chain},e} + \ln\left(\frac{\beta_{\mathrm{chain},e}}{\sinh \beta_{\mathrm{chain},e}}\right)\right], \tag{13}$$

where $\mu_B$ is the shear modulus of the elastic spring in Branch B.

For an incompressible hyperelastic material in Branch B, the Cauchy stress derived from $\psi^{(B)}$ in Eq. (13) is

$$\boldsymbol{\sigma}^{(B)} = -p^{(B)}\,\mathbf{I} + \frac{\mu_B}{3}\,\frac{\sqrt{N}}{\lambda_{\mathrm{chain},e}}\,\beta_{\mathrm{chain},e}\,\bar{\mathbf{B}}_e, \tag{14}$$

where $p^{(B)}$ is the Lagrange multiplier enforcing incompressibility for Branch B. The deviatoric Cauchy stress, which enters the flow rule below, is

$$\boldsymbol{\sigma}^{(B)}_{\mathrm{dev}} = \frac{\mu_B}{3}\,\frac{\sqrt{N}}{\lambda_{\mathrm{chain},e}}\,\beta_{\mathrm{chain},e}\,\bar{\mathbf{B}}_e^{\mathrm{dev}}, \tag{15}$$

where $\bar{\mathbf{B}}_e^{\text{dev}}$ is the deviatoric part of $\bar{\mathbf{B}}_e$. The corresponding first Piola-Kirchhoff (nominal) stress in Branch B follows

$$\mathbf{P}^{(B)} = J\,\boldsymbol{\sigma}^{(B)}\,\mathbf{F}^{-\mathrm{T}} = \boldsymbol{\sigma}^{(B)}\,\mathbf{F}^{-\mathrm{T}}, \tag{16}$$

where $J = \det \mathbf{F} = 1$.

To describe viscous flow of the nonlinear dashpot, we introduce the viscous velocity gradient $\mathbf{L}_v = \dot{\mathbf{F}}_v \mathbf{F}_v^{-1}$ and its symmetric part $\mathbf{D}_v = \text{sym}(\mathbf{L}_v)$. Consistent with the overall assumption of incompressibility, the viscous flow is taken to be isochoric, so that $\det \mathbf{F}_v = 1$ or equivalently $\text{tr}\,\mathbf{D}_v = 0$. We further introduce the nonnegative scalar flow rate $\dot{\gamma}_v$ and assume that the viscous stretching tensor is coaxial with the deviatoric Cauchy stress,

$$\mathbf{D}_v = \dot{\gamma}_v\,\mathbf{M}, \qquad \mathbf{M} = \frac{\boldsymbol{\sigma}_{\text{dev}}^{(B)}}{\|\boldsymbol{\sigma}_{\text{dev}}^{(B)}\|}, \tag{17}$$

where $\mathbf{M}$ is the unit flow direction. The corresponding equivalent shear stress in Branch B (a von Mises-type measure) is

$$\bar{\tau}^{(B)} = \left(\frac{1}{2}\boldsymbol{\sigma}_{\text{dev}}^{(B)} : \boldsymbol{\sigma}_{\text{dev}}^{(B)}\right)^{1/2}. \tag{18}$$

The polydomain LCE exhibits a rate-dependent yield-like stress that scales sub-linearly with the imposed stretch rate (Section 4), reminiscent of stress-activated viscous flow (Eyring, 1936). Inspired by this experimental observation, we adopt a power-law flow rule:

$$\dot{\gamma}_v = b\left(\bar{\tau}^{(B)}\right)^m, \tag{19}$$

where $b$ is a flow coefficient and $m$ sets the stress sensitivity of the flow rate. To offset the otherwise softening flow past yield, we further introduce a hardening-like component into $b$, which depends on the total chain stretch

$$\lambda_{\text{c}} = \sqrt{\frac{1}{3}\text{tr}\left(\mathbf{F}\mathbf{F}^{\mathrm{T}}\right)}. \tag{20}$$

In addition, we describe the temperature dependence of $b$ using a time-temperature superposition (TTS) shift factor $a_T(T)$ (Ferry and Rice, 1962; Rubinstein and Colby, 2003). The final form of $b(\lambda_{\text{c}}, T)$ is expressed as

$$b(\lambda_{\text{c}}, T) = b_0\,a_T(T)^{-m}\left[1 + \alpha\left(\lambda_{\text{c}} - 1\right)\right]^{-m}, \tag{21}$$

where $b_0$ is a reference coefficient at $T_0$ and $\alpha \geq 0$ controls the stretch-induced hardening.

The thermodynamic consistency of the model can then be checked from the reduced Clausius-Duhem inequality under the isothermal condition:

$$\mathcal{D}^{(B)} = \boldsymbol{\sigma}^{(B)} : \mathbf{D} - \dot{\psi}^{(B)} \geq 0. \tag{22}$$

Using the free energy in Eq. (13), the kinematic split in Eq. (9), and the flow rule in Eqs. (17)–(18), Eq. (22) reduces to

$$\mathcal{D}^{(B)} = \sqrt{2}\,\bar{\tau}^{(B)}\,\dot{\gamma}_v. \tag{23}$$

Therefore, when $b_0 > 0$, $m > 0$, $\alpha \geq 0$, and $a_T(T) > 0$, Eq. (19) gives $\dot{\gamma}_v \geq 0$ and hence $\mathcal{D}^{(B)} \geq 0$, satisfying the thermodynamic inequality.

### 3.4. Specialization to uniaxial loading

All experiments in this work are based on uniaxial loading at various stretch rates and temperatures. Accordingly, the deformation gradient is simplified as

$$\mathbf{F} = \text{diag}\left(\lambda^{-1/2}, \lambda^{-1/2}, \lambda\right), \tag{24}$$

where $\lambda = \lambda_3$ is the principal stretch in the loading direction. The invariant in Eq. (5) reduces to

$$\bar{I}^{(A)}(\lambda, T) = r_{\mathrm{LCE}}^{1/3}\left[\lambda^2 r_{\mathrm{LCE}}^{-1} + 2\lambda^{-1}\right]. \tag{25}$$

Substituting Eq. (25) into Eqs. (4)&(8), and eliminating the Lagrange multiplier $p^{(A)}$ via the lateral traction-free condition $P_{22}^{(A)} = 0$, we obtain the uniaxial nominal stress in Branch A as

$$P^{(A)} = \begin{cases} 0, & \lambda \le \lambda_{SL}(T) \\ \dfrac{\mu_A(T)}{(1-Q)\,\lambda}\,\dfrac{r_{\mathrm{LCE}}^{-1}\lambda^2 - \lambda^{-1}}{1-(\bar{I}^{(A)}-3)/J_m}, & \lambda > \lambda_{SL}(T) \end{cases} \tag{26}$$

where we define the temperature-dependent *soft-elastic limit* as

$$\lambda_{SL}(T) = r_{\mathrm{LCE}}(T)^{1/3}. \tag{27}$$

Eq. 26 indicates that the stress-stretch response of Branch A separates into two regimes: an ideal soft-elastic regime with zero stress ($\lambda \le \lambda_{SL}(T)$), corresponding to the collective mesogen reorientation described in Section 3.1, and a rubber-like regime with finite stress ($\lambda > \lambda_{SL}(T)$), corresponding to the aligned monodomain LCE afterward. Below $T_{ni}$ in the nematic phase, $0 < Q < 1$, $r_{\mathrm{LCE}} > 1$, and $\lambda_{SL} > 1$, leading to a finite soft-elastic regime along the stretch $\lambda$. Above $T_{ni}$ in the isotropic phase, $Q = 0$, $r_{\mathrm{LCE}} = \lambda_{SL} = 1$, and the soft-elastic regime vanishes.

For Branch B, the multiplicative decomposition in Eq. (9) reduces to

$$\lambda = \lambda_e \lambda_v, \tag{28}$$

where $\lambda_e$ and $\lambda_v$ are the elastic and viscous stretches in the spring and dashpot, respectively. With $\mathbf{F}_e = \mathrm{diag}(\lambda_e^{-1/2}, \lambda_e^{-1/2}, \lambda_e)$ from elastic incompressibility, the elastic chain stretch in Eq. (11) becomes

$$\lambda_{\mathrm{chain},e} = \sqrt{\frac{\lambda_e^2 + 2\lambda_e^{-1}}{3}}. \tag{29}$$

Substituting Eq. (29) into the full Cauchy stress in Eq. (14) and eliminating the Lagrange multiplier $p^{(B)}$ via the lateral traction-free condition $\sigma_{22}^{(B)} = 0$, we obtain the uniaxial Cauchy stress in Branch B as

$$\sigma^{(B)} = \frac{\mu_B}{3}\,\frac{\sqrt{N}}{\lambda_{\mathrm{chain},e}}\,\beta_{\mathrm{chain},e}\left(\lambda_e^2 - \lambda_e^{-1}\right). \tag{30}$$

The corresponding nominal stress follows Eq. (16) as

$$P^{(B)} = \frac{\sigma^{(B)}}{\lambda}. \tag{31}$$

The one-dimensional reduction of the flow rule in Eqs. (17)&(19) governs the evolution of the internal viscous stretch $\lambda_v$ throughout loading, unloading, and stress-free recovery.

### 3.5. Material parameter identification

We calibrate material parameters from a subset of the experimental data and use the same set to predict the remaining tests.

For the soft-elastic Branch A, we identify parameters from the single-cycle uniaxial stress-stretch curve at the lowest stretch rate, 0.001 s$^{-1}$, where the hysteresis is small, and viscoelasticity is negligible. There, we fit $\mu_{A0}$ and $J_m$ to the stress-stretch curve beyond the soft-elastic plateau, identify the soft-elastic limit $\lambda_{SL}(T)$ as the stress-free residual stretch after one cycle of loading, and calculate the order parameter $Q(T)$ from Eqs. (6) and (27). The measured values of $\lambda_{SL}(T)$ are $\lambda_{SL} = 1.8$ at 22°C, $\lambda_{SL} = 1.7$ at 40°C, and $\lambda_{SL} = 1$ at 80°C. The corresponding values of $Q(T)$ are listed in Table 1.

**Table 1**
Calibrated material parameters used in the thermoviscoelastic constitutive model. The reference temperature is $T_0 = 295$ K (22°C).

| Parameter | Physical meaning | Value |
|---|---|---|
| *Soft-elastic branch* | | |
| $\mu_{A0}$ | Reference shear modulus at $T = T_0$ | 0.0675 MPa |
| $J_m$ | Gent parameter for limiting extensibility | 4 |
| $Q$ | Nematic order parameter determining $r_{\mathrm{LCE}}(T)$ and $\lambda_{SL}(T)$ | 0.617 (22°C), 0.566 (40°C), 0 (80°C) |
| *Viscoelastic branch* | | |
| $\mu_B$ | Shear modulus of the elastic spring | 0.12 MPa |
| $m$ | Stress exponent controlling nonlinear rate dependence of viscosity | 1.425 |
| $b_0$ | Reference nonlinear flow coefficient at $T_0$ | 16 |
| $\alpha$ | Stretch hardening parameter for the nonlinear viscous term | 13 |
| $N$ | Number of Kuhn segments per chain in the eight-chain network | 1.35 |
| $a_T$ | Temperature-dependent TTS shift factor | 1 (22°C), 0.45 (40°C), 0.06 (80°C) |

For the viscoelastic Branch B, we identify parameters from the single-cycle uniaxial stress-stretch curves at other stretch rates, i.e., 0.01 $\mathrm{s}^{-1}$ and 0.1 $\mathrm{s}^{-1}$, after subtracting the fitted response of the soft-elastic Branch A. Most parameters are calibrated using the room-temperature single-cycle stress-stretch response at 22°C. The same set of parameters is then used for all temperatures, with the temperature dependence introduced primarily through the TTS shift factor $a_T(T)$.

To check the physical consistency of the calibrated $a_T(T)$, we fit it with the Williams-Landel-Ferry (WLF) equation (Williams et al., 1955; Ferry and Rice, 1962):

$$\log_{10} a_T(T) = -\frac{C_1 (T - T_0)}{C_2 + (T - T_0)}, \tag{32}$$

where $T_0 = 295$ K, $C_1 = 108.8$, and $C_2 = 5152$ K. The three calibrated values of $a_T(T)$ agree well with the WLF equation (Fig. 3).

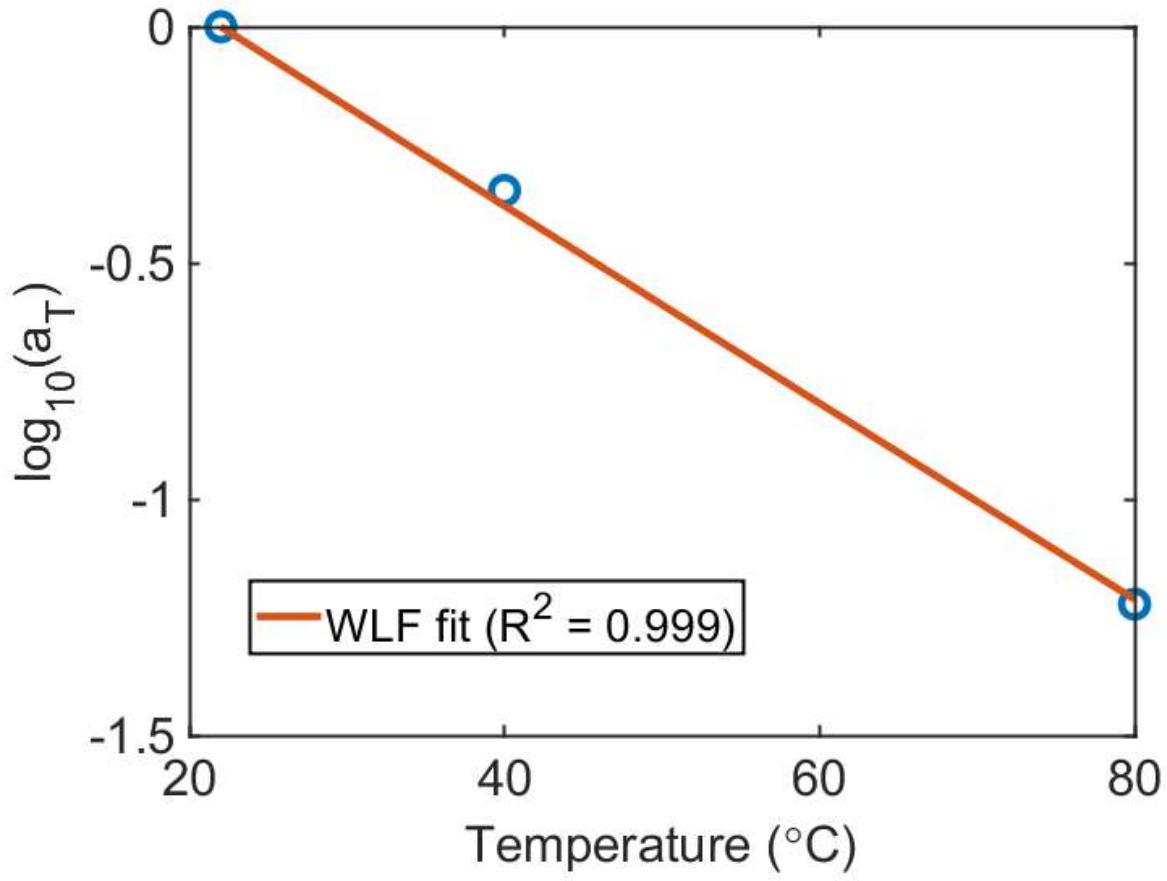

Figure 3: Calibrated time-temperature superposition shift factor $a_T(T)$ at the three test temperatures (open circles) fitted with the Williams-Landel-Ferry (WLF) equation using Eq. (32), where $T_0 = 295$ K, $C_1 = 108.8$, and $C_2 = 5152$ K.

## 4. Thermoviscoelastic single-cycle stress-stretch responses

We begin with single-cycle uniaxial stress-stretch responses of the polydomain LCE at different temperatures and stretch rates (Fig. 4). In experiments, we choose temperatures of 22, 40, and 80 °C to encompass the nematic-isotropic phase transformation ($T_{ni} \approx 60$ °C (Wei et al., 2025)): the first two lie in the nematic phase (below $T_{ni}$) and the third in the isotropic phase (above $T_{ni}$). At a prescribed temperature, we choose stretch rates of 0.1, 0.01, and 0.001 s$^{-1}$, spanning two orders of magnitude to probe the viscoelastic response. In each test, a specimen undergoes a single-cycle uniaxial loading and unloading between the undeformed state, $\lambda = 1$, and a prescribed finite stretch beyond the possible soft-elastic regime (in the nematic phase), $\lambda = 2.5$, 2.0, and 1.5, at $T = 22$, 40, and 80 °C, respectively. The experimentally measured nominal stress-stretch responses are compared with predictions of the theoretical model in Section 3 using the single set of parameters in Table 1.

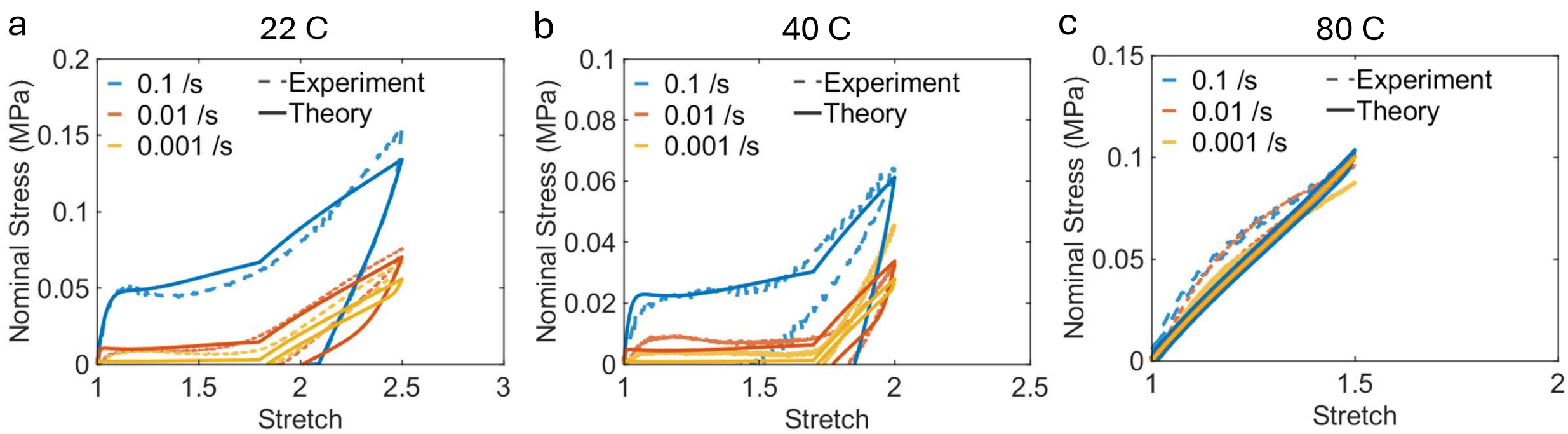

Figure 4: Single-cycle uniaxial nominal stress-stretch responses of the polydomain LCE at (a) 22°C, (b) 40°C, and (c) 80°C for stretch rates of 0.1, 0.01, and 0.001 s$^{-1}$. Dashed curves denote experimental results and solid curves denote theoretical predictions. At 22°C and 40°C, both below the nematic-isotropic transition temperature $T_{ni}$, the stress-stretch responses exhibit a pronounced soft-elastic plateau followed by stretch stiffening, all showing strong temperature and rate dependence. At 80°C above $T_{ni}$, the soft-elastic plateau disappears and the stress-stretch responses become predominantly rubber-like with substantially reduced hysteresis.

At 22 °C (Fig. 4a), across all stretch rates, the loading stress-stretch curve initially increases, reaches a plateau of soft-elastic regime with a relatively constant stress level, and subsequently increases again showing rubber-like stiffening. Unloading the sample to zero stress leads to a finite residual stretch, which will be investigated further in Sections 5&6. The single-cycle stress-stretch response depends strongly on the stretch rate: as the rate increases from 0.001 to 0.1 s$^{-1}$, stress levels of both the soft-elastic plateau and the rubber-like stiffening regimes increase by nearly an order of magnitude, the transition between the two regimes shifts left toward a smaller stretch, and the stress-stretch hysteresis increases dramatically. At the lowest rate, 0.001 s$^{-1}$, the loading and unloading curves nearly overlap, showing minimal hysteresis. In this case, dissipation from the viscoelastic branch becomes negligible, and the response of LCE is dominated by the rate-independent soft-elastic branch. During unloading, the stress returns to zero at a finite $\lambda \approx \lambda_{SL}$ rather than $\lambda = 1$, a unique feature of nematic LCE compared to conventional rubbers. In the model, we take the soft-elastic limit $\lambda_{SL} = 1.8$, equivalent to the measured finite residual stretch at 0.001 s$^{-1}$. Theoretically, during loading below $\lambda_{SL}$, the stress in the soft-elastic branch remains zero as polydomain mesogens reorient progressively toward the loading direction. Above $\lambda_{SL}$, the soft-elastic branch stiffens as a common rubber, due to the now globally aligned monodomain mesogens. At higher rates, the non-equilibrium viscoelastic branch further increases the stress level and alters the polydomain-to-monodomain transition.

At 40 °C (Fig. 4b), still below $T_{ni}$, the rate-dependent stress-stretch responses retain the same form of two regimes, but two features shift with the increase of temperature. First, the soft-elastic limit decreases to $\lambda_{SL} = 1.7$, consistent with a lower nematic order parameter $Q(T)$ and a smaller chain-anisotropy ratio $r_{\text{LCE}}(T)$ (Eqs. (6) and (27)). Second, compared to those at 22 °C, stress-stretch hysteresis decreases at each rate, a consequence of the temperature-dependent viscous resistance of the LCE, as described theoretically by the temperature-dependent shift factor $a_T(T)$ in Eq. (21) (Fig. 3). At any given rate, due to the effect of time-temperature superposition, the LCE at 40 °C responds as if it were loaded more slowly at 22 °C.

At 80 °C (Fig. 4c), above $T_{ni}$, no soft elasticity is observed in the stress-stretch responses, since the LCE transforms to the isotropic phase with $Q = 0$ and $r_{\text{LCE}} = 1$, such that Eq. (27) returns $\lambda_{SL} = 1$. The stress-stretch curves at all different rates are rubber-like with negligible rate dependence or hysteresis. The disappearance of the soft-elastic plateau above $T_{ni}$ also confirms that it arises from mesogen reorientation in the nematic phase at lower temperatures.

Fig. 4 further confirms that the proposed theoretical model successfully captures the single-cycle stress-stretch response of polydomain LCE across a range of temperatures and stretch rates. The model reproduces several characteristic features and their underlying physical mechanisms: the soft-elastic plateau arising from mesogen reorientation, the temperature-dependent soft-elastic limit $\lambda_{SL}(T)$, the post-plateau stiffening, the rate dependence of stress and hysteresis below $T_{ni}$, and the collapse to a rubber-like response above $T_{ni}$. Using the single set of parameters in Table 1, with temperature entering only through $\mu_A(T)$, $Q(T)$, and $a_T(T)$, the theoretical predictions show generally good quantitative agreement with the experimental measurements. The most noticeable discrepancy between theory and experiment appears in the unloading curve at 0.1 s$^{-1}$ and 40 °C. Although the origin of this discrepancy is not fully resolved, we attribute it to a likely larger temperature fluctuation in the Instron environmental chamber at the relatively high rate of 0.1 s$^{-1}$. Under this condition, a combination of a possibly lower actual temperature and the high stretch rate could induce additional viscoelastic-driven recovery from the unrelaxed Branch B, resulting in a smaller residual stretch below $\lambda_{SL}$, as will be detailed in Sections 5&6 and Figs. 7&10.

Moving forward, we next use the same set of parameters to investigate the temperature-dependent stress-free recovery (Section 5) and the thermoviscoelastic multi-cycle stress-stretch responses (Section 6).

## 5. Temperature-dependent stress-free recovery

The single-cycle stress-stretch responses provide a preliminary understanding of the role of soft elasticity in thermoviscoelastic behaviors of a polydomain LCE: when stretched beyond the temperature-dependent soft-elastic limit $\lambda_{SL}(T)$, the residual stretch after unloading is locked in at $\lambda_{SL}(T)$, corresponding to complete mesogen reorientation toward the loading direction. We now ask how this residual stretch evolves over prolonged stress-free conditions, whether it is permanent or transient, and how the same soft elasticity governs the time- and temperature-dependent stress-free recovery from an arbitrary prestretch, either beyond or within the soft-elastic limit $\lambda_{SL}(T)$.

To answer these questions, we propose and study a temperature-dependent stress-free recovery test (Fig. 5). In one such test, at a prescribed temperature of $T = 22$, 40, or 80 °C, we apply an instantaneous uniaxial pre-stretch $\lambda = \lambda_0$ to an LCE strip, release the load immediately, and record the evolution of recovered stretch $\lambda_{\text{rec}} = \lambda(t)$ over time (Fig. 5a). We conduct tests with a wide range of $\lambda_0$ covering values both below and above $\lambda_{SL}(T)$ at a prescribed temperature (Figs. 5b-d).

At 22 °C (Fig. 5b) and 40 °C (Fig. 5c), both in the nematic phase below $T_{ni}$, each curve of $\lambda_{\text{rec}}(t)$ starts at the prescribed $\lambda_0$, decreases sharply over the first few tens of seconds, and settles onto a plateau over a sufficiently long time (i.e., ~60 s at 22 °C and ~30 s at 40 °C). The initial decrease is generally steeper for larger $\lambda_0$ with some minor test-to-test scatter. This scatter is attributed to the manual pre-stretch and release protocol, in which the loading and unloading conditions (such as rate and holding time) are not precisely controlled. Nevertheless, we observe that the plateau of recovered stretch, $\lambda_{\text{rec}}(t \to +\infty)$, is lower at higher $T$ and further depends on the interplay between $\lambda_0$ and $\lambda_{SL}$ in a binary manner. For samples with $\lambda_0 < \lambda_{SL}$, the plateau lies below $\lambda_{SL}$ and increases with $\lambda_0$. For samples with $\lambda_0 \geq \lambda_{SL}$, the plateau converges near $\lambda_{SL}$, independent of $\lambda_0$. The converged plateau for $\lambda_0 > \lambda_{SL}$ at 22 °C settles slightly above $\lambda_{SL}$ rather than on it. The origin of this small offset is not established, but it may arise from sample-to-sample variability between the batch used for this test and the batch from which $\lambda_{SL}$ was measured in the single-cycle test. At 80 °C (Fig. 5d), in the isotropic phase above $T_{ni}$, all curves converge to $\lambda_{\text{rec}} = 1$ after approximately 20 s, regardless of $\lambda_0$, leaving no residual stretch, like a common viscoelastic rubber.

These experimental observations suggest an interplay of two mechanisms in stress-free recovery of an LCE in the nematic phase: a unified viscoelastic process from chain deformation and mesogen rotation that produces time- and temperature-dependent creep, and the soft-elastic response of the nematic network that locks in the long-term deformation as large as $\lambda_{SL}(T)$. The contribution of the latter vanishes in the isotropic phase (80 °C), such that viscoelasticity alone restores the material to its undeformed configuration. To further understand the individual contributions of these two mechanisms, we theoretically model the stress-free recovery, where the stress and stretch components in each branch are directly accessible.

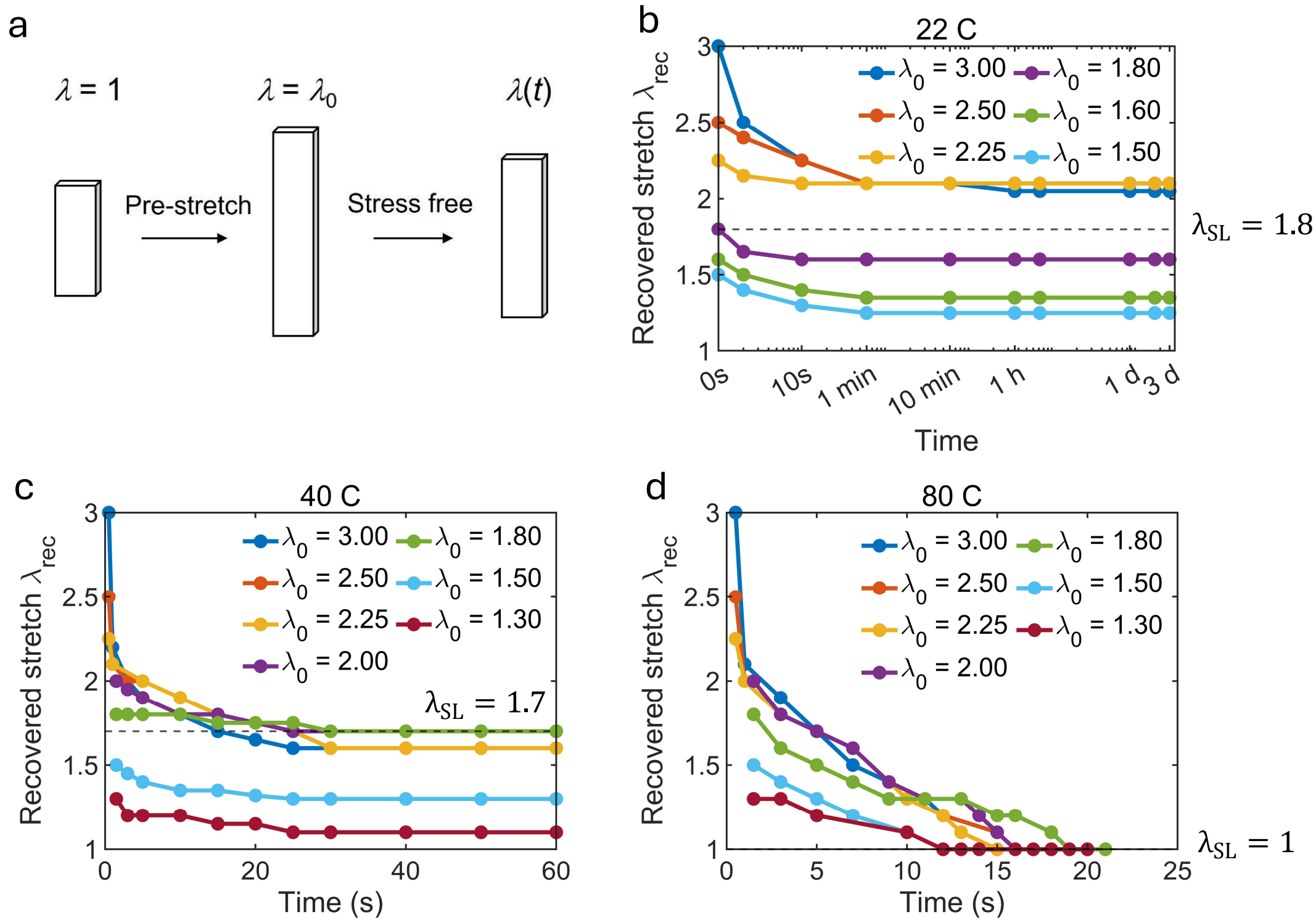

Figure 5: Temperature-dependent stress-free recovery test. (a) Schematic of the testing protocol: an initially undeformed LCE strip is instantaneously stretched to a pre-stretch $\lambda_0$, released immediately, and allowed to undergo a stress-free recovery with the time-dependent recovered stretch $\lambda_{\text{rec}} = \lambda(t)$. Throughout the test, the temperature is prescribed at $T = 22$, 40, or 80 °C. (b)–(d) Experimentally measured recovered stretch $\lambda_{\text{rec}}(t)$ for different prescribed $\lambda_0$ at 22 °C, 40 °C, and 80 °C, respectively. Solid dots represent experimental data. Solid lines serve as visual guidance. Dashed horizontal lines in (b)&(c) represent values of $\lambda_{SL}$ measured from single-cycle stress-stretch curves.

We start with a theoretical test at 22 °C with a quantitatively controlled loading path. An LCE strip is stretched to $\lambda_0 = 2.5$ at a fixed rate 0.1 s$^{-1}$, immediately unloaded at the same rate until the nominal stress reaches zero, and then held stress-free under condition $P^{(A)} + P^{(B)} = 0$ (see black curves in Figs. 6a&b as the stretch path and the corresponding total stress). Over a long time, the stretch decreases and converges to $\lambda_{\text{rec}} = \lambda_{\text{SL}} = 1.8$, consistent with the experimental observation.

Using the theoretical model, we next decompose the stress-stretch responses into the two branches and examine their individual contributions (Fig. 6b blue and orange curves). The soft-elastic Branch A begins with an ideal zero-stress soft-elastic plateau up to $\lambda = \lambda_{\text{SL}} = 1.8$, beyond which the mesogens are fully aligned to form a monodomain and the branch exhibits rubber-like stiffening with increasing stress. Upon unloading, this branch retraces its loading path until it reaches and remains at the stress-free state $\lambda = \lambda_{\text{SL}} = 1.8$. The viscoelastic Branch B, by contrast, carries a finite stress from the onset of loading, representing the rate-dependent viscoelastic resistance of the LCE as a whole. During unloading at the fixed rate 0.1 s$^{-1}$, the total stress reaches zero while the total stretch has not yet reached $\lambda_{SL}$, due to the nonequilibrium state of the viscoelastic branch. At this point, the soft-elastic Branch A still carries a tensile stress while the viscoelastic Branch B carries a compressive stress of equal magnitude. During the subsequent stress-free recovery, viscoelasticity in Branch B drives $P^{(B)}$ toward zero, while elasticity in Branch A drives the material to re-enter the soft-elastic regime at $\lambda = \lambda_{\text{SL}} = 1.8$. Eventually, both branches reach equilibrium with zero stress (Figs. 6b&c) and a permanent recovered stretch of $\lambda_{\text{rec}} = \lambda_{\text{SL}} = 1.8$. The soft elastic Branch A thus dictates the long-term equilibrium recovery stretch through its nematic spring, while the viscoelastic Branch B governs the recovery kinetics through the nonlinear viscous flow.

We further confirm this physical picture at the kinematic level by plotting the viscous stretch $\lambda_v$ and the elastic stretch $\lambda_e$ of the viscoelastic Branch B (Fig. 6a, orange and yellow curves), where the total stretch satisfies $\lambda = \lambda_e \lambda_v$. Throughout the test, the viscous stretch $\lambda_v$ lags behind the total stretch $\lambda$. Consequently, $\lambda_v$ falls below $\lambda$ during loading, rises above $\lambda$ during unloading, and eventually converges to $\lambda = \lambda_{\mathrm{SL}} = 1.8$ in the long term. Correspondingly, the elastic stretch $\lambda_e$ is initially above unity (tensile), decreases upon unloading, drops slightly below unity (compressive) due to incomplete relaxation of the viscous dashpot, and finally returns to unity in the long term.

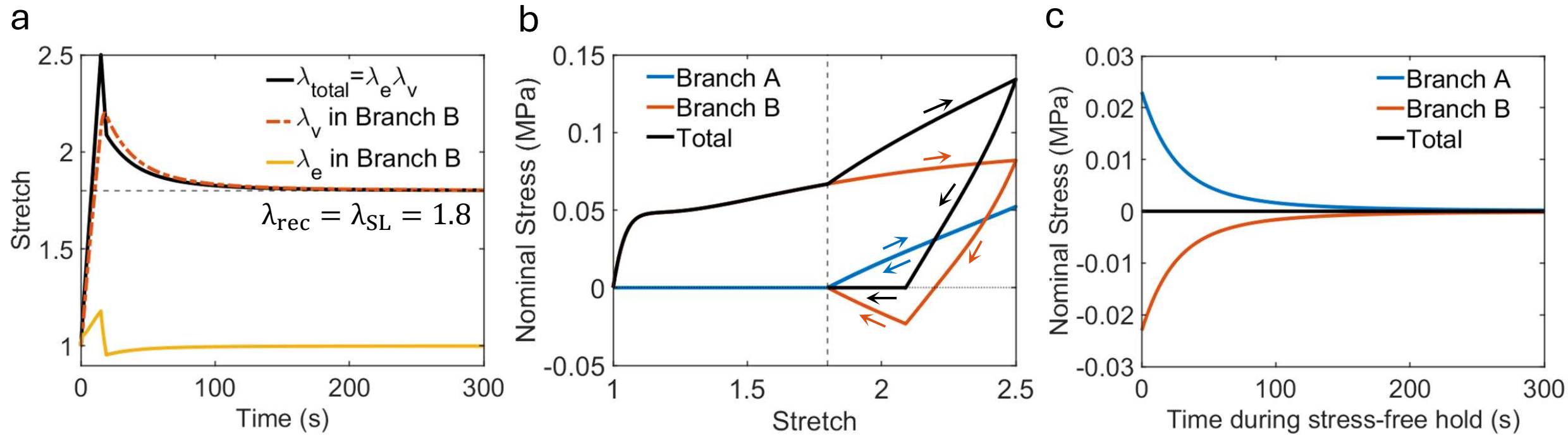

Figure 6: Theoretical model of stress-free recovery at 22 °C. An LCE strip is stretched to $\lambda_0 = 2.5$ at a fixed rate 0.1 $s^{-1}$, immediately unloaded at the same rate until the nominal stress reaches zero, and then held stress-free. (a) Time evolution of the total stretch $\lambda = \lambda_e \lambda_v$, along with the viscous stretch $\lambda_v$ and the elastic stretch $\lambda_e$ of the viscoelastic Branch B, through loading, unloading, and stress-free hold. The long-term equilibrium recovered stretch approaches $\lambda_{\mathrm{rec}} = \lambda_{SL} = 1.8$. (b) Full stress-stretch response histories of the soft-elastic Branch A, the viscoelastic Branch B, and the total LCE. Arrows indicate the direction of evolution. (c) Time evolution of the nominal stresses in Branches A and B during stress-free hold, satisfying $P^{(A)} + P^{(B)} = 0$.

We next extend this theoretical study across a range of pre-stretch $\lambda_0$, stretch rates, and temperatures (Fig. 7). To systematically analyze the coupled effects of these factors, we generate heat maps of the long-term equilibrium recovered stretch $\lambda_{\mathrm{rec}}(t \to +\infty)$ as a function of $\lambda_0$ and stretch rate at 22 °C, 40 °C, and 80 °C (Fig. 7a). At 22 and 40 °C, both in the nematic phase below $T_{ni}$, the heat maps show two regimes separated by a rate- and temperature-dependent boundary (white dashed line), where the equilibrium $\lambda_{\mathrm{rec}}$ changes with $\lambda_0$ and the stretch rate on the left but remains constant as $\lambda_{SL}$ on the right. At 80 °C, in the isotropic phase above $T_{ni}$, the heat map collapses to a uniform recovered stretch $\lambda_{\mathrm{rec}} = 1$. These heat maps are replotted in Fig. 7b as the equilibrium $\lambda_{\mathrm{rec}}$ versus $\lambda_0$ at different rates and temperatures, and compared with the experimental results. Despite the uncontrolled loading and unloading rates in experiments, the theoretical model captures the overall trend across all three temperatures with reasonable quantitative agreement.

At low stretch rates (0.001 and 0.01 $s^{-1}$) in the nematic phase (22 °C and 40 °C), the viscoelastic branch is nearly fully relaxed throughout loading, unloading, and stress-free hold, such that the soft elastic Branch A alone sets $\lambda_{\mathrm{rec}}$. When $\lambda_0 < \lambda_{SL}$, the pre-stretch falls within the soft-elastic regime and stores no elastic energy in Branch A, leaving no restoring stress to drive recovery. Consequently, $\lambda_{\mathrm{rec}} \approx \lambda_0$, following a diagonal dotted line in Fig. 7b. The soft-elastic state is effectively locked in, producing a pseudo-plastic response despite the absence of any internal damage (to be validated in Fig. 11). When $\lambda_0 \geq \lambda_{SL}$, as discussed above, mesogen reorientation is complete at $\lambda_{SL}$ and the rubber-like elasticity of Branch A drives the LCE back to $\lambda_{\mathrm{rec}} \approx \lambda_{SL}$ regardless of $\lambda_0$ (the horizontal dashed line in Fig. 7b). The recovered stretch in the low-rate limit therefore collapses to $\lambda_{\mathrm{rec}} \approx \min(\lambda_0, \lambda_{SL}(T))$, reflecting the interplay between the pre-stretch $\lambda_0$ and the soft-elastic limit $\lambda_{SL}$.

Finite stretch rates (0.1 and 0.5 $s^{-1}$) introduce an additional viscoelastic-driven recovery from the unrelaxed Branch B. At these rates, the viscous stretch $\lambda_v$ in Branch B lags behind the total stretch $\lambda$ during loading, producing an elastic stretch $\lambda_e > 1$ (Fig. 6a) that stores elastic energy in the spring of Branch B. During the stress-free hold, the resulting restoring stress from the spring pulls the equilibrium $\lambda_{\mathrm{rec}}$ below the low-rate limit (blue and orange lines in Fig. 7b), with lower $\lambda_{\mathrm{rec}}$ at higher rates. At all rates, the maximum $\lambda_{\mathrm{rec}}$ remains set by $\lambda_{SL}$, but higher rates require larger $\lambda_0$ to overcome the additional restoring stress from Branch B to reach $\lambda_{\mathrm{rec}} = \lambda_{SL}$. As the temperature increases from 22 °C to 40 °C, $\lambda_{SL}$ decreases and the linear regime $\lambda_{\mathrm{rec}} \approx \lambda_0$ in Fig. 7b shifts to the left. Moreover, viscoelastic relaxation accelerates with increasing temperature, bringing the finite-rate curves closer to the low-rate limit. Finally, at 80 °C above $T_{ni}$, $\lambda_{SL} = 1$ and $\lambda_{\mathrm{rec}} = 1$ at all rates: the LCE behaves as an elastic rubber.

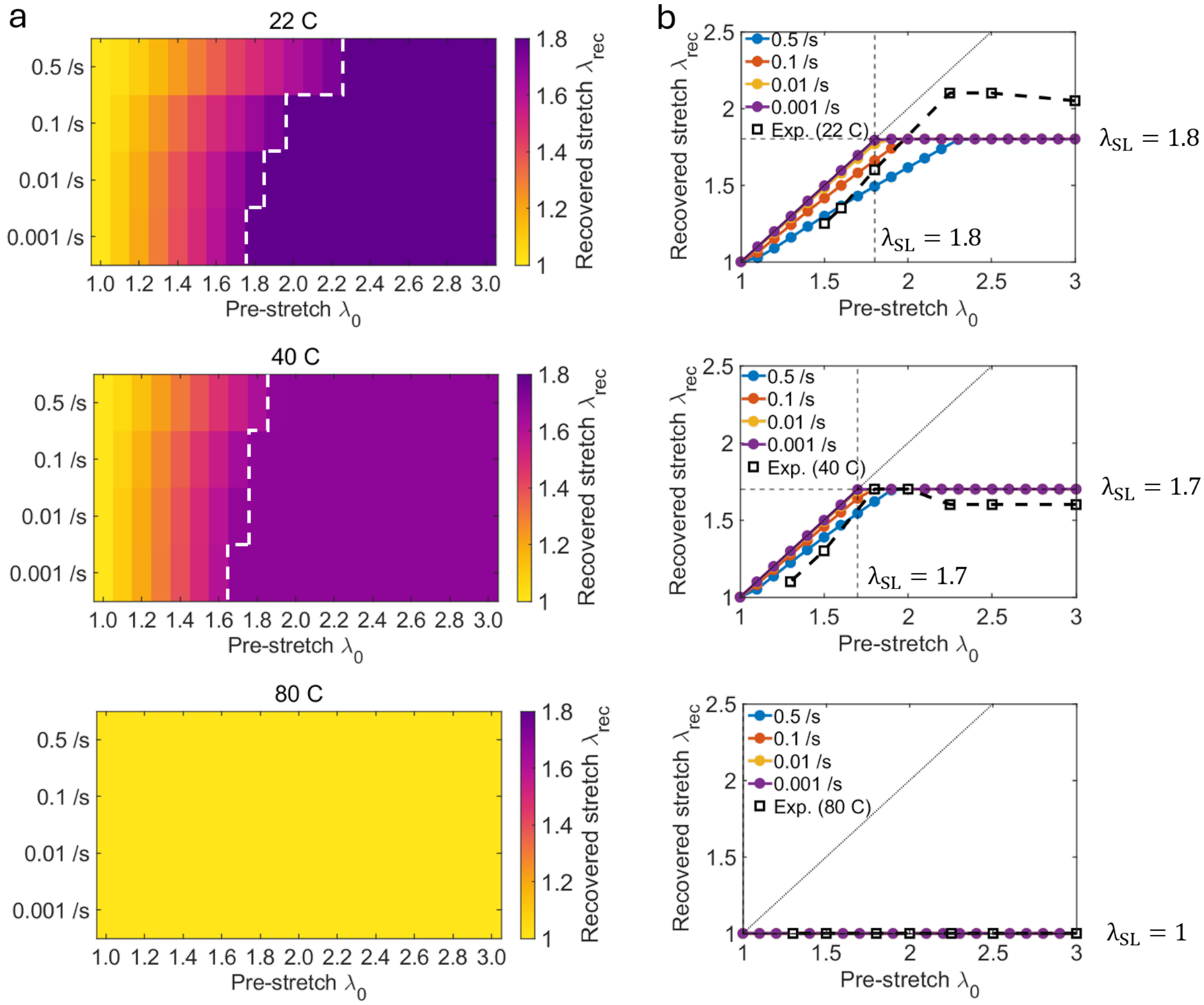

Figure 7: Theoretically predicted long-term equilibrium recovered stretch $\lambda_{\text{rec}}$ across a range of pre-stretch $\lambda_0$, stretch rates, and temperatures. (a) Heat maps of long-term equilibrium $\lambda_{\text{rec}}$ as a function of $\lambda_0$ and stretch rate at 22 °C, 40 °C, and 80 °C. The white dashed line at 22°C and 40°C separates the heat map into two regimes, where the equilibrium $\lambda_{\text{rec}}$ changes with $\lambda_0$ and the stretch rate on the left but remains constant as $\lambda_{SL}$ on the right. (b) Corresponding plots of the theoretically predicted equilibrium $\lambda_{\text{rec}}$ versus $\lambda_0$ at different rates and temperatures. Open squares represent experimental results from Fig. 5. Diagonal dotted lines denote $\lambda_{\text{rec}} = \lambda_0$. Horizontal and vertical dashed lines mark $\lambda_{SL}$ at the corresponding temperature. At 80°C, $\lambda_{SL} = 1$ and both theoretical and experimental results show $\lambda_{\text{rec}} = 1$ at all rates.

## 6. Thermoviscoelastic multi-cycle stress-stretch responses

We have so far studied how temperature-dependent soft elasticity regulates the thermoviscoelastic behavior of the polydomain LCE in the short and long terms after a single loading cycle. We now ask how the same soft elasticity regulates the behavior under repeated cyclic loads with no stress-free hold, and whether our theoretical model with the same set of parameters can capture the multi-cycle response.

In experiments, we impose a multi-cycle protocol at a prescribed constant stretch rate during both loading and unloading. The protocol progressively increases the applied maximum stretch, denoted $\lambda_{\text{applied}}$, over cycles. The protocol terminates each cycle when the nominal stress first returns to zero during unloading, records the cycle-wise residual stretch $\lambda_{\text{res}}$, and starts a new cycle immediately afterward (Fig. 1d).

Fig. 8 compares experimental results and theoretical predictions of cyclic stress-stretch responses at two temperatures (22 and 40 °C, both in the nematic phase below $T_{ni}$) and two stretch rates (0.01 and 0.001 s$^{-1}$). Each plot overlays the multi-cycle stress-stretch response with the corresponding single-cycle response from Fig. 4. During the multi-cycle test, loading in each subsequent cycle initiates from the zero-stress residual stretch $\lambda_{\mathrm{res}}$ of the previous cycle, retraces the preceding unloading path upward with minor stiffening due to viscoelastic relaxation, and then resumes the same stress-stretch response as if it were continuous loading beyond the applied maximum stretch $\lambda_{\mathrm{applied}}$ from the previous cycle. The theoretical predictions show perfect overlapping of single-cycle and multi-cycle stress-stretch curves at all temperatures and rates. The experimental measurements show some discrepancies between single-cycle and multi-cycle curves, which we attribute to sample-to-sample variability across synthesized batches. Overall, the theory reproduces both the single-cycle and multi-cycle responses with good qualitative agreement across all cases and without any additional cyclic-specific fitting.

From both experimental and theoretical results in Fig. 8, we extract the immediate residual stretch $\lambda_{\mathrm{res}}$ in each cycle and plot it against the applied maximum stretch $\lambda_{\mathrm{applied}}$ (Fig. 9). In the low-rate limit (0.001 $\mathrm{s}^{-1}$), the residual stretch $\lambda_{\mathrm{res}}$ follows the same trend observed in the stress-free recovery test (Fig. 7): $\lambda_{\mathrm{res}} \approx \min(\lambda_{\mathrm{applied}}, \lambda_{SL}(T))$. That is, $\lambda_{\mathrm{res}} \approx \lambda_{\mathrm{applied}}$ for $\lambda_{\mathrm{applied}} < \lambda_{SL}$ and saturates at $\lambda_{SL}$ for $\lambda_{\mathrm{applied}} \geq \lambda_{SL}$. At the higher rate (0.01 $\mathrm{s}^{-1}$) and the same temperature, $\lambda_{\mathrm{res}}$ deviates in opposite directions on the two sides of $\lambda_{SL}$: it falls below $\lambda_{\mathrm{applied}}$ (the diagonal dotted line) for $\lambda_{\mathrm{applied}} < \lambda_{SL}$ and rises above $\lambda_{SL}$ (the horizontal dashed line) for $\lambda_{\mathrm{applied}} > \lambda_{SL}$. This rate-dependent deviation diminishes as temperature increases (40 °C, Fig. 9b). The theoretical predictions capture both the overall trend and these more subtle rate-dependent features on both sides of $\lambda_{SL}$, with relatively good quantitative agreement.

We extend the theoretical modeling to a wider range of $\lambda_{\mathrm{applied}}$ and stretch rate beyond those accessible in experiments. After each cycle, we increase $\lambda_{\mathrm{applied}}$ by 0.1 up to $\lambda_{\mathrm{applied}} = 3$, corresponding to 20 cycles in total. Analogous to Fig. 7, we generate heat maps of the cycle-wise $\lambda_{\mathrm{res}}$ as a function of $\lambda_{\mathrm{applied}}$ and stretch rate at 22 °C and 40 °C (Fig. 10a), and replot these data as $\lambda_{\mathrm{res}}$ versus $\lambda_{\mathrm{applied}}$ at different rates and temperatures in Fig. 10b.

In all cases in Fig. 10b, the overall trend and the opposite deviation of higher-rate curves from the low-rate limit (0.001 $\mathrm{s}^{-1}$) on the two sides of $\lambda_{SL}$ remain consistent with Fig. 9. With the increase of $\lambda_{\mathrm{applied}}$ and greater number of total cycles, curves at lower rates (0.001 and 0.01 $\mathrm{s}^{-1}$) at both 22 °C and 40 °C gradually settle onto a rate-dependent steady-state plateau that coincides with or lies above $\lambda_{SL}(T)$. A lower stretch rate and a higher temperature lead to a lower plateau and an earlier onset stretch for the plateau. Consequently, within the range $\lambda_{\mathrm{applied}} \leq 3$ studied here, the 0.1 $\mathrm{s}^{-1}$ curve never reaches steady state at 22 °C and reaches it near $\lambda_{\mathrm{applied}} \approx 3$ at 40 °C, while the 0.5 $\mathrm{s}^{-1}$ curve does not reach steady state at either temperature. We attribute both the rate-dependent deviation from the low-rate limit and the rate-dependent steady-state plateau to the cumulative, cycle-by-cycle contribution of the viscous stretch $\lambda_v$ in Branch B: because the cyclic loading includes no stress-free dwell, the history-dependent $\lambda_v$ accumulates and carries over to each subsequent cycle, causing $\lambda_{\mathrm{res}}$ to drift continuously from the low-rate limit until reaching steady state.

The rate dependence also produces crossovers between a curve that has reached steady state (e.g., 0.01 $\mathrm{s}^{-1}$ at 22 °C) and ones that have not (e.g., 0.5 $\mathrm{s}^{-1}$ at 22 °C). Before the crossover, $\lambda_{\mathrm{res}}$ is smaller at higher rates and lower temperatures, consistent with the additional viscoelastic-driven recovery from the unrelaxed Branch B discussed in Section 5 (stress-free recovery). As discussed, this effect is also a likely source of the noticeable discrepancy between theory and experiment in the single-cycle unloading curve at 0.1 $\mathrm{s}^{-1}$ and 40 °C (Fig. 4b).

The recovered stretch $\lambda_{\mathrm{rec}}$ in stress-free recovery (Fig. 7) and the residual stretch $\lambda_{\mathrm{res}}$ in multi-cycle tests (Fig. 10) share the same low-rate limit, $\min(\lambda_0, \lambda_{SL}(T))$ or $\min(\lambda_{\mathrm{applied}}, \lambda_{SL}(T))$, respectively. At higher rates, both $\lambda_{\mathrm{rec}}$ and $\lambda_{\mathrm{res}}$ fall below this low-rate limit when the applied stretch ($\lambda_0$ or $\lambda_{\mathrm{applied}}$) is smaller than $\lambda_{SL}(T)$, manifesting the additional viscoelastic-driven recovery that drives the LCE further into the soft-elastic regime below $\lambda_{SL}$. $\lambda_{\mathrm{rec}}$ and $\lambda_{\mathrm{res}}$ diverge, however, as the applied stretch increases beyond $\lambda_{SL}$: $\lambda_{\mathrm{rec}}$ converges to $\lambda_{SL}$ because the stress-free hold allows the material to reach long-term equilibrium, whereas $\lambda_{\mathrm{res}}$ converges to a higher plateau because the unrelaxed viscous stretch accumulates over successive cycles without a stress-free dwell.

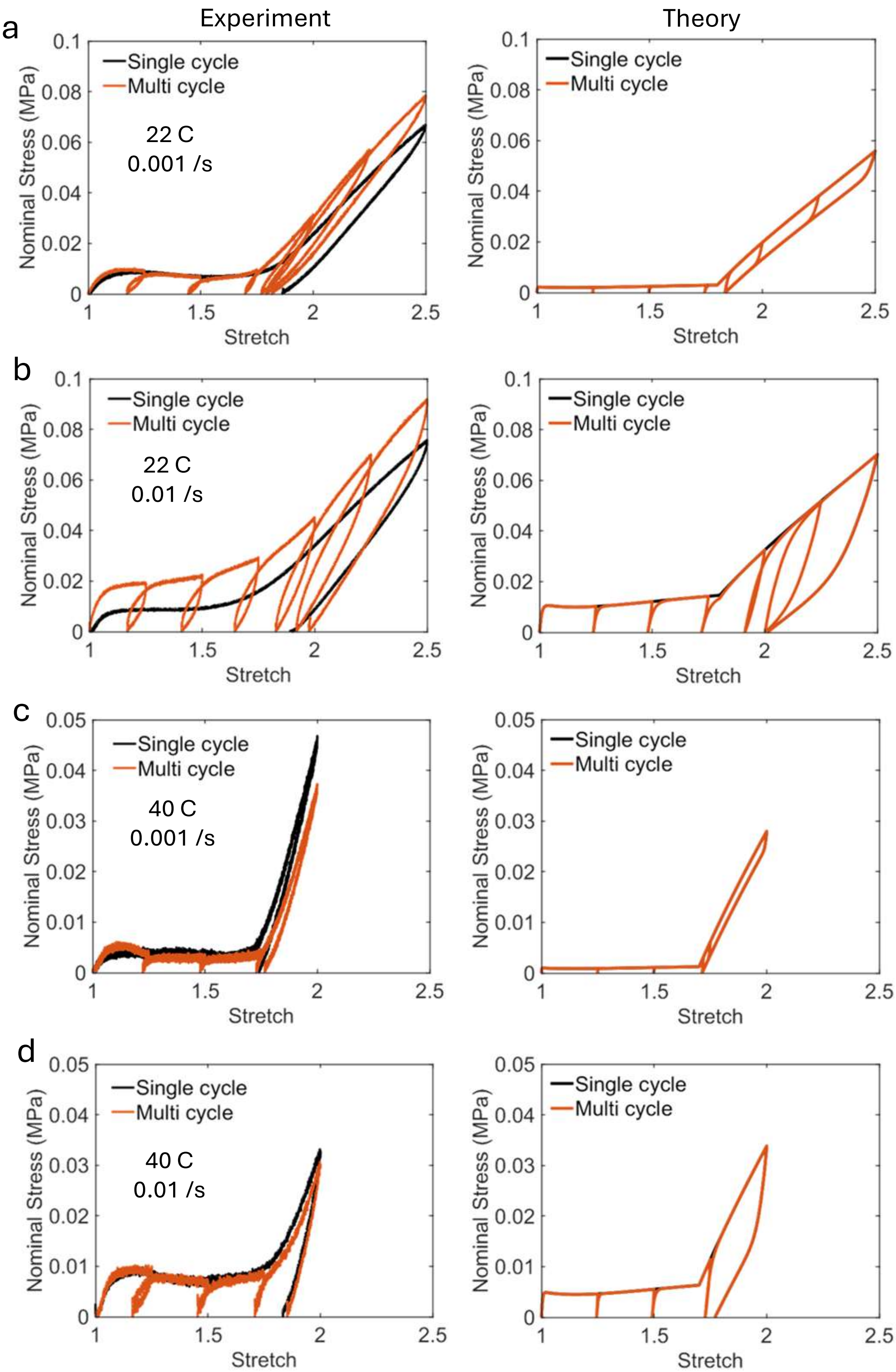

Figure 8: Comparison of experimental results (left column) and theoretical predictions (right column) of multi-cycle stress-stretch responses. Each plot overlays the multi-cycle stress-stretch response with the corresponding single-cycle response from Fig. 4. (a) 22 °C and 0.001 $s^{-1}$. (b) 22 °C and 0.01 $s^{-1}$. (c) 40 °C and 0.001 $s^{-1}$. (b) 40 °C and 0.01 $s^{-1}$.

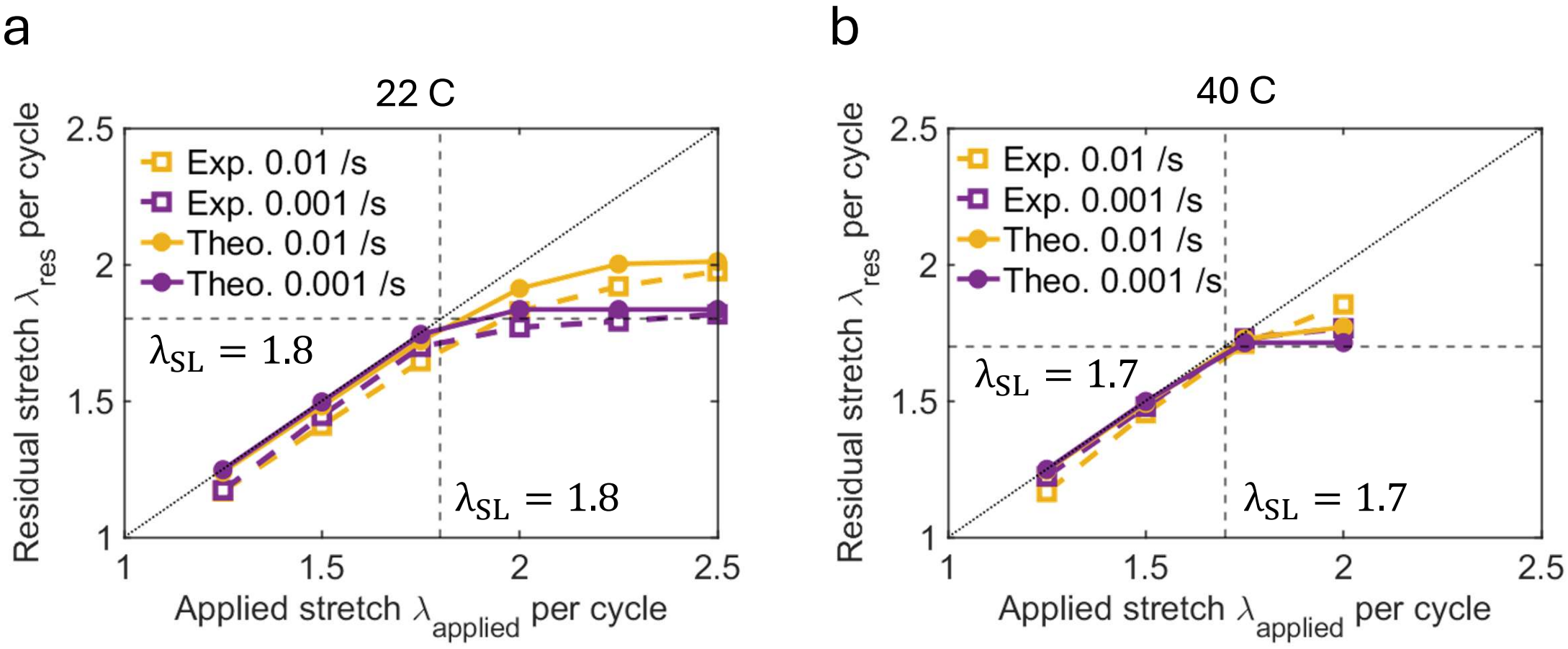

Figure 9: Residual stretch $\lambda_{\text{res}}$ in each cycle from Fig. 8 as a function of the applied maximum stretch $\lambda_{\text{applied}}$. Both theoretical and experimental results are shown at (a) 22°C and (b) 40°C for stretch rates of 0.01 and 0.001 $s^{-1}$. Diagonal dotted lines denote $\lambda_{\text{res}} = \lambda_{\text{applied}}$. Horizontal and vertical dashed lines mark $\lambda_{SL}$ at the corresponding temperature.

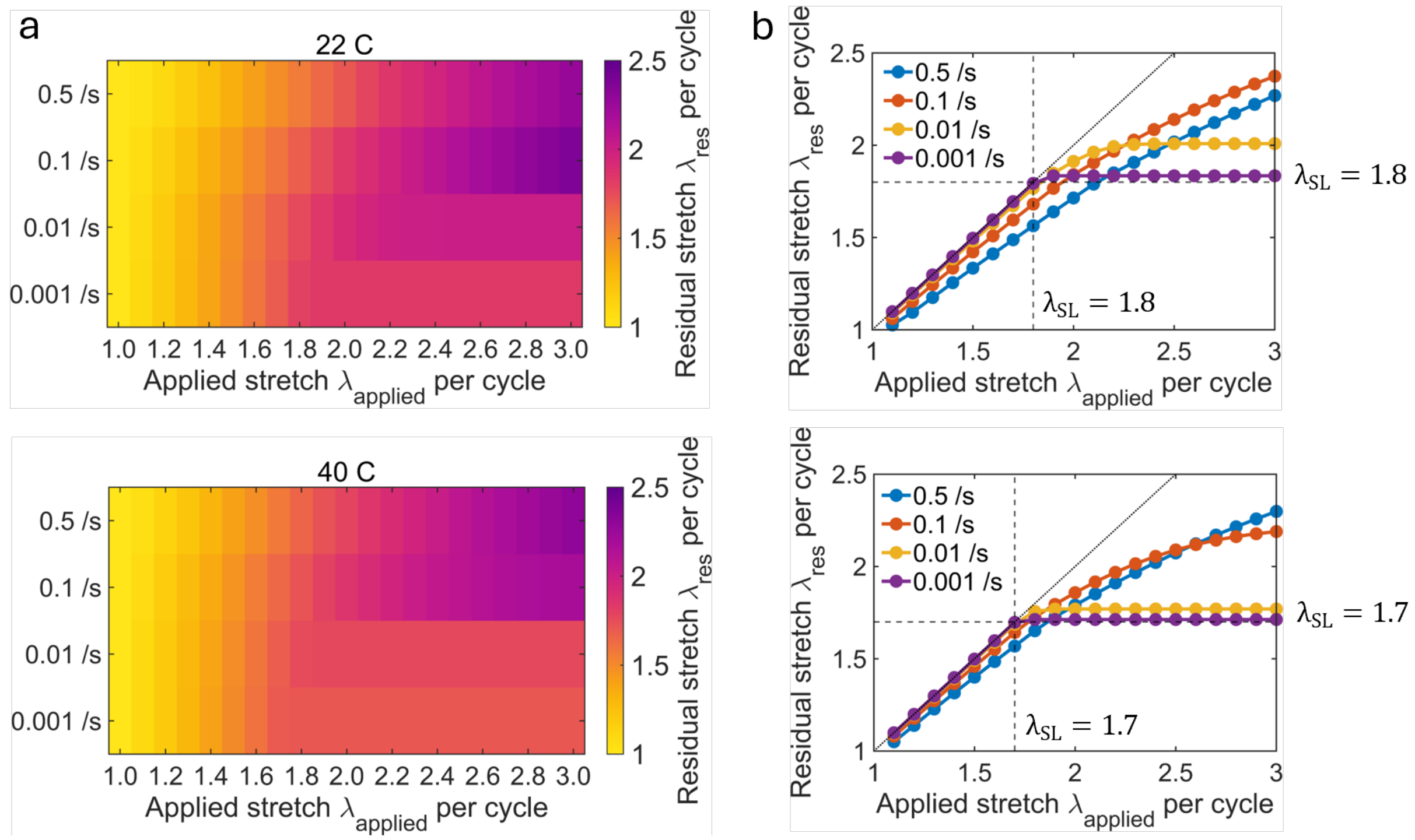

Figure 10: Theoretically predicted residual stretch $\lambda_{\text{res}}$ in each cycle across a range of applied maximum stretch $\lambda_{\text{applied}}$ and stretch rates beyond those accessible in experiments, at 22 °C and 40 °C. (a) Heat maps of $\lambda_{\text{res}}$ as a function of $\lambda_{\text{applied}}$ and stretch rate. (b) Corresponding plots of the theoretically predicted $\lambda_{\text{res}}$ versus $\lambda_{\text{applied}}$ at different rates. Diagonal dotted lines denote $\lambda_{\text{res}} = \lambda_{\text{applied}}$. Horizontal and vertical dashed lines mark $\lambda_{SL}$ at the corresponding temperature.

Finally, we rule out irreversible internal damage as a factor in our analysis. Experimentally, we subject an LCE strip to multi-cycle or single-cycle loads up to $\lambda_{\text{applied}} = 3$ at 0.01 $s^{-1}$ and 22 °C, unload the sample, immediately heat it to 80 °C (isotropic phase), cool it back to 22 °C in a water reservoir, and apply the same cyclic-load protocol again (Fig. 11a). The stress-stretch responses before and after heat treatment overlap nearly perfectly under both multi-cycle and single-cycle loading conditions (Figs. 11b&c). Heating the polydomain LCE above $T_{ni}$ transforms the material into the isotropic phase with $\lambda_{SL} = 1$, erasing any residual stretch or internal memory imparted by non-equilibrium viscoelastic relaxation and soft elasticity. Subsequent cooling therefore restores the LCE to its virgin polydomain configuration and original viscoelastic properties. The near-perfect overlap of responses before and after heat treatment confirms that the mechanical loading induces no irreversible internal damage.

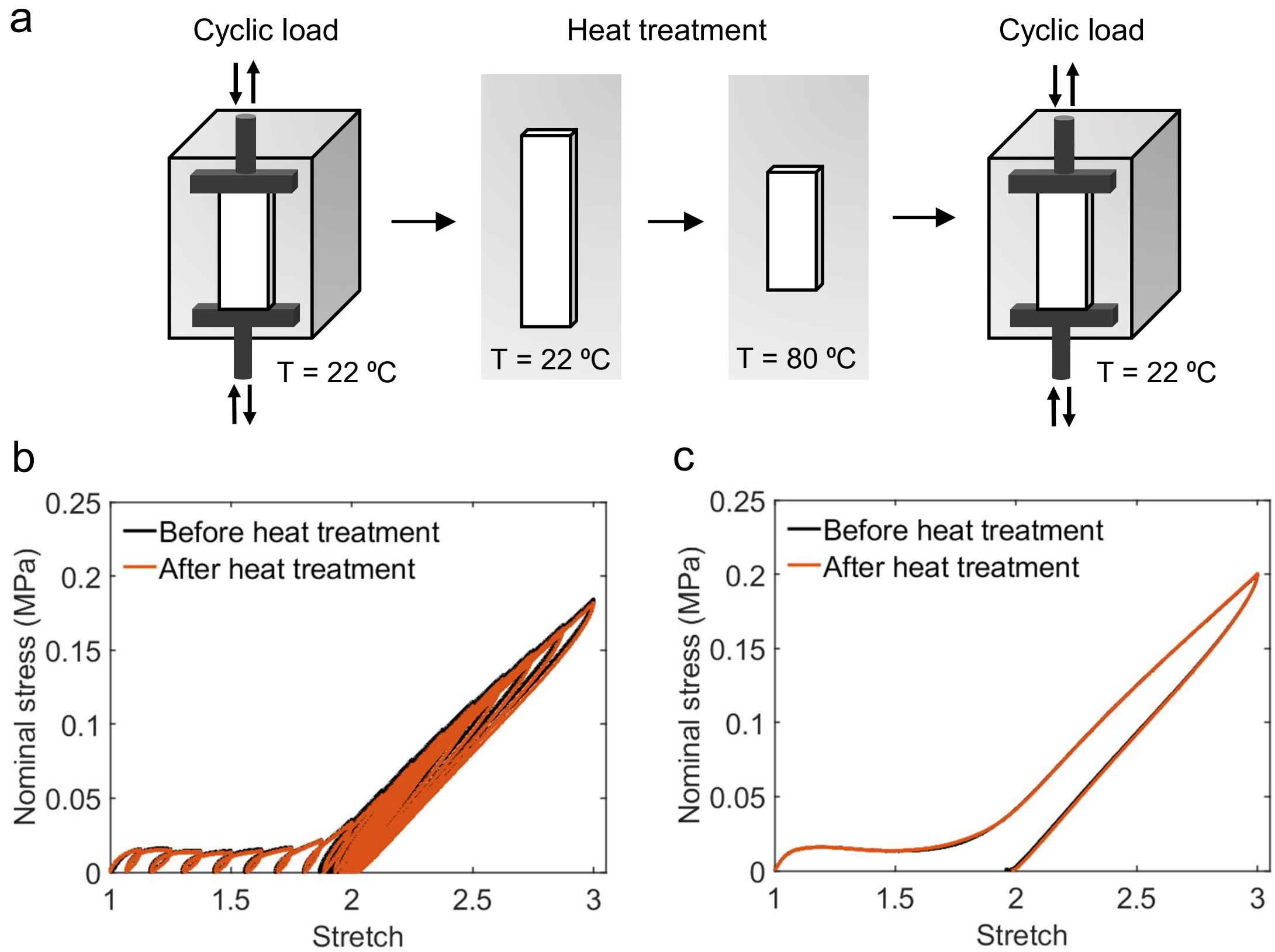

Figure 11: Thermal recovery test of polydomain LCE that rules out irreversible internal damage. (a) Schematic of the protocol: an LCE strip is subjected to multi-cycle or single-cycle loads up to $\lambda_{\text{applied}} = 3$ at 0.01 s$^{-1}$ and 22 °C, unloaded, immediately heated to 80 °C (isotropic phase), cooled back to 22 °C, and subjected to the same cyclic-load protocol again. The stress-stretch responses are measured before and after heat treatment under both (b) multi-cycle and (c) single-cycle loading conditions.

## 7. Discussion and conclusion

We have investigated the fundamental role of soft elasticity in the path-dependent, finite-deformation thermoviscoelastic response of polydomain LCEs across a range of stretch rates and temperatures, using both experiments and theoretical modeling. Three loading protocols are examined in detail, including single-cycle loading, stress-free recovery, and multi-cycle loading. The theoretical framework incorporates two mechanisms that jointly govern the path-dependent thermoviscoelasticity of polydomain LCEs. A rate-independent soft-elastic Branch A captures the ideal soft-elastic response driven by mesogen reorientation through a temperature-dependent plateau that terminates at the soft-elastic limit $\lambda_{SL}(T)$, with $\lambda_{SL}(T) > 1$ in the nematic phase and $\lambda_{SL} = 1$ in the isotropic phase above the nematic-isotropic transition temperature $T_{ni}$. A viscoelastic Branch B captures the unified time- and temperature-dependent relaxation of the LCE network through a Maxwell-like model comprising a nonlinear elastic spring and a nonlinear viscous dashpot in series. Calibrated with a single set of parameters, the model quantitatively reproduces the experimental results across all three protocols over a large range of rates and temperatures, including the single- and multi-cycle stress-stretch responses, the non-equilibrium and equilibrium recovered stretch in stress-free recovery, and the cycle-wise residual stretch under cyclic loads.

The detailed interplay between soft elasticity driven by mesogen reorientation in Branch A and viscoelasticity in Branch B is revealed through further analysis of the experimental and theoretical results across different protocols. The temperature-dependent soft-elastic limit $\lambda_{SL}(T)$ emerges as the central governing quantity shared by all three protocols: it defines the low-rate-limit behavior as a backbone, while the time- and temperature-dependent viscoelasticity controls the deviation from this backbone at finite rates. Consequently, compared to stress-free recovery which allows full relaxation over a long term, the multi-cycle responses exhibit a steady-state residual stretch that lies above $\lambda_{SL}$ over many cycles due to the accumulated unrelaxed viscous stretch. A thermal recovery test with heat treatment above $T_{ni}$ fully erases any residual stretch or internal memory, restores the LCE to its virgin configuration and properties, and confirms that no irreversible internal damage is induced throughout loading and unloading.

The close integration of experiment and theory for path-dependent, finite-deformation thermoviscoelastic behavior of LCEs has received comparatively little attention in previous studies, particularly in the context of cyclic loading and the associated cycle-wise accumulation of residual deformation. Yet cyclic behavior and properties govern the accuracy, repeatability, and stability of LCE-based applications that involve long-term use, such as architected and functionally graded LCE structures and thermal- or light-driven LCE actuators. The combined experimental and theoretical framework presented here thus provides a predictive basis for the design and control of thermoviscoelastic LCEs under diverse mechanical and thermal histories.

Several idealizations in the present theoretical framework point to directions for future work. The current model lumps all dissipative mechanisms into a single viscoelastic branch. Future extensions could resolve the separate contributions of polymer-network viscoelasticity, sub-domain reorganization, and other dissipative processes in polydomain LCEs under different loading conditions. The soft-elastic branch assumes ideal soft elasticity with a zero plateau stress, whereas real LCEs typically exhibit a finite onset stress and gradual softening. Incorporating semi-soft elasticity is therefore a natural refinement. The current formulation is also effectively homogenized at the polydomain scale and does not resolve the orientation distribution of individual nematic sub-domains, which could be explicitly tracked in future mesoscale or director-resolved extensions. The protocols and analysis presented here have focused on uniaxial loading, while the model can be readily extended to more complex and practically relevant scenarios such as history-dependent multi-axial loading. Finally, a finite element implementation of the theoretical framework would be a valuable next step toward quantitative simulation and design of LCE components and devices for engineering applications.

## Acknowledgments

This work was supported by the National Science Foundation through grants CMMI-2146409 and CMMI-2440758.

## Conflict of interest

The authors declare no conflict of interest.

# CRediT authorship contribution statement

**Zhengxuan Wei:** Conceptualization, Methodology, Experiments, Visualization, Writing – original draft. **Beijun Shen:** Conceptualization, Methodology, Model, Investigation, Visualization, Writing – original draft, Writing – review & editing. **Zumrat Usmanova:** Model, Visualization, Writing – review & editing. **Umme Hani Bootwala:** Experiments. **Ruobing Bai:** Conceptualization, Methodology, Supervision, Writing – review & editing.

# References

Aharoni, H., Xia, Y., Zhang, X., Kamien, R., Yang, S., 2018. Universal inverse design of surfaces with thin nematic elastomer sheets. Proceedings of the National Academy of Sciences 115, 7206–7211.

Ahmadi, A., Maghsoodi, N., 2024. Creasing instability of polydomain nematic elastomers in compression. Journal of the Mechanics and Physics of Solids 193, 105870.

Amin, A., Lion, A., Sekita, S., Okui, Y., 2006. Nonlinear dependence of viscosity in modeling the rate-dependent response of natural and high damping rubbers in compression and shear: Experimental identification and numerical verification. International journal of plasticity 22, 1610–1657.

Annapooranan, R., Yeerella, R., Chambers, R., Li, C., Cai, S., 2024. Soft elasticity enabled adhesion enhancement of liquid crystal elastomers on rough surfaces. Proceedings of the National Academy of Sciences 121, e2412635121.

Arruda, E.M., Boyce, M.C., 1993. A three-dimensional constitutive model for the large stretch behavior of rubber elastic materials. Journal of the Mechanics and Physics of Solids 41, 389–412.

Azoug, A., Vasconcellos, V., Dooling, J., Saed, M., Yakacki, C., Nguyen, T., 2016. Viscoelasticity of the polydomain-monodomain transition in main-chain liquid crystal elastomers. Polymer 98, 165–171.

Bai, R., Bhattacharya, K., 2020. Photomechanical coupling in photoactive nematic elastomers. Journal of the Mechanics and Physics of Solids 144.

Bergström, J.S., Boyce, M.C., 1998. Constitutive modeling of the large strain time-dependent behavior of elastomers. Journal of the Mechanics and Physics of Solids 46, 931–954.

Biggins, J., Warner, M., Bhattacharya, K., 2009. Supersoft elasticity in polydomain nematic elastomers. Physical Review Letters 103, 037802.

Biggins, J., Warner, M., Bhattacharya, K., 2012. Elasticity of polydomain liquid crystal elastomers. Journal of the Mechanics and Physics of Solids 60, 573–590.

Bladon, P., Terentjev, E.M., Warner, M., 1993. Transitions and instabilities in liquid crystal elastomers. Physical Review E 47, R3838.

Chehade, A.E.H., Shen, B., Yakacki, C.M., Nguyen, T.D., Govindjee, S., 2024. Finite element modeling of viscoelastic liquid crystal elastomers. International Journal for Numerical Methods in Engineering 125, e7510.

Chen, B., Liu, C., Xu, Z., Wang, Z., Xiao, R., 2024a. Modeling the thermo-responsive behaviors of polydomain and monodomain nematic liquid crystal elastomers. Mechanics of Materials 188, 104838.

Chen, B., Yang, H., Xiao, R., 2025. Experimental characterization and modeling anisotropic mechanical responses of liquid crystal elastomers with exchangeable disulfide bonds. Extreme Mechanics Letters , 102435.

Chen, B., Zhu, L., Yang, H., Xiao, R., 2026. A thermo-order-mechanical coupled model for main-chain isotropic-genesis polydomain liquid crystal elastomers. Mechanics of Materials 212, 105531.

Chen, W., Tong, D., Meng, L., Tan, B., Lan, R., Zhang, Q., Yang, H., Wang, C., Liu, K., 2024b. Knotted artificial muscles for bio-mimetic actuation under deepwater. Advanced Materials 36, 2400763.

Choi, S., Guo, H., Kim, B., Seo, J.H., Terentjev, E., Saed, M., Ahn, S.k., 2025. Harnessing extreme internal damping in polyrotaxane-incorporated liquid crystal elastomers for pressure-sensitive adhesives. Advanced Functional Materials 35, 2413824.

Chung, C., Luo, C., Yakacki, C.M., Song, B., Long, K., Yu, K., 2024. Revealing the unusual rate-dependent mechanical behaviors of nematic liquid crystal elastomers. International Journal of Solids and Structures 292, 112712.

Conti, S., DeSimone, A., Dolzmann, G., 2002a. Semisoft elasticity and director reorientation in stretched sheets of nematic elastomers. Physical Review E 66, 061710.

Conti, S., DeSimone, A., Dolzmann, G., 2002b. Soft elastic response of stretched sheets of nematic elastomers: a numerical study. Journal of the Mechanics and Physics of Solids 50, 1431–1451.

Dai, Z., Wen, Y., Chen, Z., Chen, Y., Yang, Y., Gao, M., Chen, Y., Xu, F., 2025. Unusual stretching–twisting of liquid crystal elastomer bilayers. Journal of the Mechanics and Physics of Solids 198, 106066.

DeSimone, A., Dolzmann, G., 2002. Macroscopic response of nematic elastomers via relaxation of a class of so(3)-invariant energies. Archive for Rational Mechanics and Analysis 161, 181 – 204.

Eyring, H., 1936. Viscosity, plasticity, and diffusion as examples of absolute reaction rates. The Journal of Chemical Physics 4, 283–291.

Ferry, J., Rice, S., 1962. Viscoelastic Properties of Polymers. American Institute of Physics.

Fu, K., Zhao, Z., Jin, L., 2019. Programmable granular metamaterials for reusable energy absorption. Advanced Functional Materials 29, 1901258.

Guin, T., Settle, M., Kowalski, B., Auguste, A., Beblo, R., Reich, G., White, T., 2018. Layered liquid crystal elastomer actuators. Nature Communications 9, 2531.

He, Q., Zheng, Y., Wang, Z., He, X., Cai, S., 2020. Anomalous inflation of a nematic balloon. Journal of the Mechanics and Physics of Solids 142, 104013.

Hertlein, N., Shen, B., Kang, S.H., Buskohl, P.R., 2023. Computational analysis of buckling-based stiffness programming of liquid crystal elastomer beams, in: Proceedings of the ASME 2023 International Design Engineering Technical Conferences and Computers and Information in Engineering Conference, ASME. p. V008T08A040.

Hotta, A., Terentjev, E., 2003. Dynamic soft elasticity in monodomain nematic elastomers. The European Physical Journal E 10, 291–301.

Jeon, S.Y., Shen, B., Traugutt, N.A., Zhu, Z., Fang, L., Yakacki, C.M., Nguyen, T.D., Kang, S.H., 2022. Synergistic energy absorption mechanisms of architected liquid crystal elastomers. Advanced Materials 34, 2200272.

Jiang, Y., Xu, Z., Xiao, R., Govindjee, S., Nguyen, T.D., 2026. A viscoelastic micro-stretch theory for monodomain nematic liquid crystal elastomers. Journal of the Mechanics and Physics of Solids 207, 106412.

Koshimizu, N., Saed, M., 2025. Switchable pressure-sensitive adhesion in nematic side-chain liquid crystal elastomers. Macromolecules 58, 12191–12200.

Kularatne, R., Kim, H., Boothby, J., Ware, T., 2017. Liquid crystal elastomer actuators: Synthesis, alignment, and applications. Journal of Polymer Science Part B: Polymer Physics 55, 395–411.

Kutsyy, A., Wihardja, A., Lee, V., Bhattacharya, K., 2025. Characterization of the soft behavior of nematic elastomers over a range of temperature and strain rates. arXiv preprint arXiv:2512.05146 .

Lee, V., Wihardja, A., Bhattacharya, K., 2023. A macroscopic constitutive relation for isotropic-genesis, polydomain liquid crystal elastomers. Journal of the Mechanics and Physics of Solids 179, 105369.

Li, Y., Luo, C., Yu, K., Wang, X., 2021. Remotely controlled, reversible, on-demand assembly and reconfiguration of 3D mesostructures via liquid crystal elastomer platforms. ACS Applied Materials & Interfaces 13, 8929–8939.

Li, Y., Teixeira, Y., Parlato, G., Grace, J., Wang, F., Huey, B., Wang, X., 2022. Three-dimensional thermochromic liquid crystal elastomer structures with reversible shape-morphing and color-changing capabilities for soft robotics. Soft Matter 18, 6857–6867.

Liang, X., Li, D., 2022. A programmable liquid crystal elastomer metamaterials with soft elasticity. Frontiers in Robotics and AI 9, 849516.

Linares, C., Traugutt, N., Saed, M., Linares, A., Yakacki, C., Nguyen, T., 2020. The effect of alignment on the rate-dependent behavior of a main-chain liquid crystal elastomer. Soft Matter 16, 8782–8798.

Liu, K., Hacker, F., Daraio, C., 2021. Robotic surfaces with reversible, spatiotemporal control for shape morphing and object manipulation. Science Robotics 6, eabf5116.

Luo, C., Chung, C., Yakacki, C., Long, K., Yu, K., 2021. Real-time alignment and reorientation of polymer chains in liquid crystal elastomers. ACS Applied Materials & Interfaces 14, 1961–1972.

Maghsoodi, A., Saed, M.O., Terentjev, E.M., Bhattacharya, K., 2023. Softening of the hertz indentation contact in nematic elastomers. Extreme Mechanics Letters 63, 102060.

Rubinstein, M., Colby, R.H., 2003. Polymer Physics. Oxford University Press.

Saed, M., Gablier, A., Terentjev, E., 2022. Exchangeable liquid crystalline elastomers and their applications. Chemical Reviews 122, 4927–4945.

Saed, M., Torbati, A., Nair, D., Yakacki, C., 2016. Synthesis of programmable main-chain liquid-crystalline elastomers using a two-stage thiol-acrylate reaction. JoVE (Journal of Visualized Experiments) , e53546.

Shen, B., 2025. Viscoelastic Behavior and Energy Absorption of Main-Chain Liquid Crystal Elastomers-Based Architected Materials. Ph.D. Dissertation. Johns Hopkins University. Baltimore, MD, USA.

Shen, B., Jiang, Y., Yakacki, C.M., Kang, S.H., Nguyen, T.D., 2026. Combining stretching-dominated and bending-dominated dissipation behavior to optimize energy absorption in liquid crystal elastomer-based lattice structures. Journal of the Mechanics and Physics of Solids 209, 106497.

Sydney Gladman, A., Matsumoto, E., Nuzzo, R., Mahadevan, L., Lewis, J., 2016. Biomimetic 4D printing. Nature Materials 15, 413–418.

Tokumoto, H., Zhou, H., Takebe, A., Kamitani, K., Kojio, K., Takahara, A., Bhattacharya, K., Urayama, K., 2021. Probing the in-plane liquid-like behavior of liquid crystal elastomers. Science Advances 7, eabe9495.

Traugutt, N., Mistry, D., Luo, C., Yu, K., Ge, Q., Yakacki, C., 2020. Liquid-crystal-elastomer-based dissipative structures by digital light processing 3D printing. Advanced Materials 32, 2000797.

Traugutt, N., Volpe, R., Bollinger, M., Saed, M., Torbati, A., Yu, K., Dadivanyan, N., Yakacki, C., 2017. Liquid-crystal order during synthesis affects main-chain liquid-crystal elastomer behavior. Soft Matter 13, 7013–7025.

Ula, S., Traugutt, N., Volpe, R., Patel, R., Yu, K., Yakacki, C., 2018. Liquid crystal elastomers: an introduction and review of emerging technologies. Liquid Crystals Reviews 6, 78–107.

Urayama, K., Kohmon, E., Kojima, M., Takigawa, T., 2009. Polydomain–monodomain transition of randomly disordered nematic elastomers with different cross-linking histories. Macromolecules 42, 4084–4089.

Usmanova, Z., Bai, R., 2026. Anomalous thermomechanical actuation of liquid crystal elastomer balloons. Extreme Mechanics Letters 83, 102448.

Verwey, G., Warner, M., 1997. Compositional fluctuations and semisoftness in nematic elastomers. Macromolecules 30, 4189–4195.

Wang, X., Han, J., Xu, H., Ji, H., Yue, Z., Zhang, R., Li, B., Ji, Y., Li, Z., Wang, P., Lu, T.J., 2024. Nonlinear mechanical behaviour and visco-hyperelastic constitutive description of isotropic-genesis, polydomain liquid crystal elastomers at high strain rates. Journal of the Mechanics and Physics of Solids 193, 105882.

Wang, Z., El Hajj Chehade, A., Govindjee, S., Nguyen, T., 2022. A nonlinear viscoelasticity theory for nematic liquid crystal elastomers. Journal of the Mechanics and Physics of Solids 163, 104829.

Ware, T., McConney, M., Wie, J., Tondiglia, V., White, T., 2015. Voxelated liquid crystal elastomers. Science 347, 982–984.

Warner, M., Terentjev, E., 2007. Liquid Crystal Elastomers. Oxford University Press.

Wei, C., Cao, S., Zhou, Y., Lin, D., Jin, L., 2023a. Rate-dependent stress-order coupling in main-chain liquid crystal elastomers. Soft Matter 19, 7923–7936.

Wei, C., Zhou, Y., Hsu, B., Jin, L., 2024. Exceptional stress-director coupling at the crack tip of a liquid crystal elastomer. Journal of the Mechanics and Physics of Solids 183, 105522.

Wei, Z., Bootwala, U., Bai, R., 2025. Synthesis-processing-property relationships in thermomechanics of liquid crystal elastomers. Journal of the Mechanics and Physics of Solids 196, 105977.

Wei, Z., Wang, P., Bai, R., 2023b. Thermomechanical coupling in polydomain liquid crystal elastomers. Journal of Applied Mechanics 91, 021001.

White, T., Broer, D., 2015. Programmable and adaptive mechanics with liquid crystal polymer networks and elastomers. Nature Materials 14, 1087–1098.

Wihardja, A., Nieto-Fuentes, J.C., Rittel, D., Bhattacharya, K., 2026. High strain rate behavior of liquid crystal elastomers. Journal of the Mechanics and Physics of Solids 209, 106501.

Williams, M.L., Landel, R.F., Ferry, J.D., 1955. The temperature dependence of relaxation mechanisms in amorphous polymers and other glass-forming liquids. Journal of the American Chemical Society 77, 3701–3707.

Wu, J., Yao, S., Zhang, H., Man, W., Bai, Z., Zhang, F., Wang, X., Fang, D., Zhang, Y., 2021. Liquid crystal elastomer metamaterials with giant biaxial thermal shrinkage for enhancing skin regeneration. Advanced Materials 33, 2106175.

Yakacki, C., Saed, M., Nair, D., Gong, T., Reed, S., Bowman, C., 2015. Tailorable and programmable liquid-crystalline elastomers using a two-stage thiol–acrylate reaction. RSC Advances 5, 18997–19001.

Zhou, Y., Wei, C., Jin, L., 2025. A modified semi-soft model of liquid crystal elastomers: Application to elastic and viscoelastic responses. Journal of the Mechanics and Physics of Solids 196, 106027.